%% file: main.tex
\documentclass[sigplan]{acmart}

\renewcommand\footnotetextcopyrightpermission[1]{} 
\acmISBN{} 
\acmDOI{} 
\startPage{1}

\setcopyright{none}

\usepackage{booktabs}   
\usepackage{subcaption} 

\usepackage[normalem]{ulem} 
\usepackage{listings}

\definecolor{blue}{RGB}{38,139,210}
\definecolor{cyan}{RGB}{42,161,152}
\lstdefinelanguage{Fireiron}
{
        morekeywords=[1]{
        to, tile, load, distributedStore, split, MatMul, Move, done, val, Swizzle, Fluent,
        epilog, unrollTile, sync, threadLayout, layout, toColMajor, noSync, toRowMajor, Strided, swizzle, unrollSplit,
        Cst, HexCst, reuseBuffer, rowMapping, colMapping, pad, String, unroll, Matrix},
morekeywords=[2]{
            Kernel, Block,Warp, Lane, Thread, GL, SH, RF, FR, RowMajor, ColMajor, Float, float, FP32, FP16},
morecomment=[l]{//},
}

\lstdefinestyle{C}{
morekeywords={syncthreads, __syncthreads},
}
\lstdefinestyle{fireiron}{
    keywordstyle=\bfseries\color{blue}, 
    keywordstyle= [2]{\bfseries\color{cyan}},
	basicstyle=\ttfamily\scriptsize, 
    commentstyle=\itshape\color{black!60}, 
	stringstyle=\itshape, 
	numbers=none, 
	numberstyle=\scriptsize, 
	stepnumber=1, 
    tabsize = 2,
	numbersep=8pt, 
	showstringspaces=false, 
	breaklines=true, 
	backgroundcolor = \color{black!10},
	frame=lines, 
	abovecaptionskip=.5\baselineskip, 
	aboveskip=\baselineskip,
	escapechar=@,
	captionpos=b,
	mathescape=true
}

\begin{document}

\title[Fireiron: High-Performance Linear Algbra on GPUs]{Fireiron: A Scheduling Language for High-Performance Linear Algebra on GPUs}         
\thanks{$^{\dagger}$This work was done while the authors were at NVIDIA}%


\author{Bastian Hagedorn$^{ \dagger}$}
\affiliation{
  \institution{University of M\"unster}            
}
\email{b.hagedorn@wwu.de}          

\author{Archibald Samuel Elliott$^{ \dagger}$}
\affiliation{
  \institution{lowRISC}            
}
\email{sam@lenary.co.uk}          

\author{Henrik Barthels$^{ \dagger}$}
\affiliation{
  \institution{AICES, RWTH Aachen University}            
}
\email{barthels@aices.rwth-aachen.de}          

\author{Rastislav Bodik$^{ \dagger}$}
\affiliation{
  \institution{University of Washington}            
}
\email{bodik@cs.washington.edu}          

\author{Vinod Grover}
\affiliation{
  \institution{NVIDIA}            
}
\email{vgrover@nvidia.com}          

\begin{abstract}
\input{tex/abstract}
\end{abstract}

\ccsdesc[500]{Software and its engineering~General programming languages}
\ccsdesc[300]{Social and professional topics~History of programming languages}


\maketitle


\section{Introduction}
\input{tex/intro}


\section{Motivation}
\input{tex/goals}

\section{Specs and Decompositions}
\input{tex/specs}

\section{Expressing Advanced Optimizations}
\input{tex/optimization}

\section{Evaluation}
\input{tex/evaluation}

\section{Related Work}
\input{tex/related}

\section{Conclusion}
\input{tex/conclusion}

%
%


\bibliography{bibfile}

%

\end{document}

%% file: tex/abstract.tex
Achieving high-performance GPU kernels requires optimizing algorithm implementations to the targeted GPU architecture.
It is of utmost importance to fully use the compute and memory hierarchy, as well as available specialised hardware.

Currently, vendor libraries like cuBLAS and cuDNN provide the best performing implementations of GPU algorithms.
However the task of the library programmer is incredibly challenging:
for each provided algorithm, high-performance implementations have to be developed for all commonly used architectures, input sizes, and different storage formats.
These implementations are generally provided as optimized assembly code because performance-critical architectural features are only exposed at this level.
This prevents reuse between different implementations of even the same algorithm, as simple differences can have major effects on low-level implementation details.


In this paper we introduce Fireiron, a DSL and compiler which allows the specification of high-performance GPU implementations as compositions of simple and reusable building blocks.
We show how to use Fireiron to optimize matrix multiplication implementations, achieving performance matching hand-coded CUDA kernels, even when using specialised hardware such as NIVIDA Tensor Cores, and outperforming state-of-the-art implementations provided by cuBLAS by more than 2$\times$.

%% file: tex/intro.tex
Implementing high-performance kernels for GPUs is incredibly challenging.
Achieving performance close to the theoretical peak requires the careful application of a plethora of optimizations to optimally use the compute and memory hierarchy of the target architecture.
In order to maximize available memory bandwidth, one needs to make use of vectorized loads and stores, padding to avoid shared memory bank conflicts, or storage layout transformations.
Optimizations affecting the compute hierarchy include warp specialization, swizzling, or the exploitation of special built-in instructions like NVIDIA's WMMA (warp-level matrix multiply accumulate).

Figuring out which optimizations lead to the best performance is a tedious task.
Today, kernels are either written in low-level programming languages such as CUDA or PTX-assembly, or they are generated using high-level frameworks such as Halide~\cite{DBLP:journals/tog/Ragan-KelleyAPLAD12,DBLP:conf/pldi/Ragan-KelleyBAPDA13} or TVM~\cite{DBLP:conf/osdi/ChenMJZYSCWHCGK18}.
In the first case, low-level implementation details are interwoven with and obscure the intended functionality.
This makes it hard to try out different optimizations as it usually requires rewriting large parts of the existing code.
To mitigate this, high-level frameworks such as Halide separate the description of the computation (algorithm) from its optimizations (schedule) using a separate \emph{scheduling language} that describes the optimizations to apply.
Here, however, users are still limited to the set of optimizations and application the specific framework supports.

In this paper, we propose Fireiron, a framework for gradual forging of high-performance GPU kernels.
Fireiron is a language and compiler which, similar to Halide and TVM, provides a small set of \emph{decompositions} which a programmer uses to design implementations of \emph{specifications}.
In Fireiron, important implementation concerns such as the mapping of data to threads are first-class concepts of the language and are specified before less important decisions are made.
This allows, for instance, for a programmer to decide how matrices are stored across the GPU, with the remainder of the implementation following these prior constraints.

Similar to Halide, a \emph{specification} represents the computation to implement (\emph{algorithms} in Halide) and a \emph{decomposition} describes how to (partially) implement a given specification (\emph{schedule} in Halide).
Our decompositions are composable and modular, so programmers can start an implementation with a simple decomposition and then gradually \emph{refine} it in order to specialize it for the current target hardware.
This modularity allows decompositions to be reused when targeting different architectures.

When implementing high-performance applications, GPU programmers usually rely extensively on libraries of fine-tuned routines such as cuBLAS and cuDNN.
One limitation is that inputs to these routines must be passed in addressible storage such as global memory or shared memory.
This prevents programmers from reusing values stored in, for example registers or FMA argument collectors, whose address cannot be taken.
This means existing implementations cannot be decomposed into smaller, flexible components from which new, efficient implementations could be composed by library users.

Consider the case when the input matrices are distributed across the registers of the threads of a cooperative thread array (CTA).
Ideally, we want to specify the location of the matrix (in registers) when calling the library procedure.
However, today, the procedure needs to be compiled for a particular size of the matrix and a particular number of threads in the CTA.
Additionally, the implementation of the producers of the matrices needs to be coordinated so that the matrices are placed in the registers expected by the consumers.
This coordination prevents independent development and reuse of implementations.
Fireiron is a language and compiler architecture that allows efficient library procedures to be broken into reusable pieces.

Since every DSL is only as good as the abstractions it provides, experts often struggle to break out of restrictions in situations where they need to make use of optimizations for which suitable abstractions cannot be defined easily.
In contrast to existing high-level frameworks like Halide, Fireiron allows the programmer to regain full control and provide custom implementations (e.g., handwritten inline assembly), for arbitrary specifications at any point in a decomposition.
This allows performance experts to fine-tune specific parts of their implementations using hand-crafted solutions to specifications.

To summarize, Fireiron is a framework for simplifying the development of high-performance GPU kernels, targeted towards experts.
We achieve this by achieving a balance between high-level abstractions and control, as well as by providing novel mechanisms to faciliate reuse of optimized implementations that go beyond reusing tuned library routines.
In this paper, we make the following contributions:
\begin{enumerate}
    \item We introduce Fireiron, a high-level language and compiler providing easy-to-use \emph{Decompositions} to implement \emph{Specifications} of GPU computations.
    \item We enable rapid prototyping of complex optimizations by introducing \emph{Refinements} for decompositions which allow gradual improvement of an existing implementation with minimal effort.
    We illustrate these concepts using the example of matrix multiplication.
    \item We show how to use and extend Fireiron's basic decompositions to make use of the TensorCores in newer NVIDIA architectures, and show how to define and reuse optimizations as first-class objects in Fireiron.
    \item We evaluate our approach by comparing against hand-written CUDA code targeting different architectures.
            We express the same optimizations in Fireiron and show that our generated code achieves the same performance as the hand-tuned kernels.
            Additionally we compare our approach to cuBLAS and show that by using Fireiron, in combination with \texttt{mma.sync} assembly instructions and carefully chosen tile sizes, we are able to outperform high-performance library implementations by up to 2x.
\end{enumerate}

%% file: tex/goals.tex
Fireiron's design is based upon four main goals, and the mechanisms we used to achieve them.

\subsection{\emph{Control}: Fine-grained Level of Control over the Implementation Strategy.}
Every domain-specific language is only as good as the abstractions it provides to its user.
When striving for optimal performance, there will always be implementations for which no suitable abstractions are available in high-level DSLs.
In order to avoid fighting restrictive abstractions, Fireiron offers kernel programmers the ability to regain full control at any point and for any part of the implementation.
By allowing the programmer to insert their own specialized code as \emph{micro-kernels} for matching specifications, we achieve a balance between leveraging high-level abstractions and maximizing productivity and performance.

\paragraph{Mechanism:}
The programmer specifies the implementation strategy using a decomposition language.
This language controls the decomposition of the computation, the placement of data into memory, the communication of data between levels of the memory hierarchy, as well as the mapping of computation onto the compute elements.
Decomposing a specification yields a new sub-specification which describes the problem left to implement.
A programmer decomposes a specification until it is \emph{executable}, that is it matches either the specification of a built-in instruction such as fused-multiply-add (FMA), or the specification of a user-provided micro-kernel for which they have provided an implementation.
During code generation we then either emit the built-in instruction or the provided code snippet.

\subsection{\emph{Reusability}: Reuse of Implementation Decompositions.}
We offer the kernel programmer the reuse of previously developed high-performance decompositions, including those that result in generated code which typically cannot be implemented as a traditional library procedure.

\paragraph{Mechanism:}
The implementation of a kernel is hierarchical and each fragment of the implementation can be described with a concise specification.
Specifications are implemented with decompositions that are stored as first-class objects and can therefore be reused.
Essentially, a decomposition breaks existing efficient library procedures (such as a kernel-level GEMM computation) into reusable pieces which can be stored as first-class objects.
This allows the extraction and reuse of specific parts of an efficient implementation such as a tiled warp-level GEMM whose operands are stored in shared memory or registers.

In more detail, the implementation is a decomposition tree whose leaves are specifications (representing the remaining sub-computations to implement).
In a \emph{final} implementation, the leaf specifications are executable, e.g. built-in instructions such as FMA or user-provided micro-kernels.
In a partial implementation, the leaves can be specifications that still need to be implemented.

%

\subsection{\emph{Flexibility}: Target (Specialized) Instructions with Arbitrary Granularity.}
The architecture of GPUs changes rapidly and significantly.
For example, the recent NVIDIA Volta and Turing architectures contain specialized MMA (Matrix-Multiply Accumulate) units, so-called TensorCores, which are able to efficiently compute matrix multiplications using small groups of cooperating threads.
Currently, this functionality is exposed in two ways: as 8-thread HMMA instructions in PTX (\texttt{mma.sync}), and as warp-wide WMMA instructions in CUDA.
The decomposition language must be able to express implementations which make use of these specialized instructions despite the varying granularity of their cooperating threads.

\paragraph{Mechanism:}
In Fireiron, kernels, micro-kernels, and built-in instructions are described using the same concept: \emph{specifications}.
Once a programmer decomposes a problem into a specification that is executable, i.e., a specification for which we know how to generate code, they can decide to stop decomposing the specification and instead pass the implementation to the code generator.

Imagine we decompose a kernel-level M$\times$N$\times$K-matrix multiplication kernel to a warp-level 16$\times$16$\times$16 matrix multiplication computation.
We can either decompose this specification further into several thread-level FMA instructions or stop decomposing here and generate a single WMMA invocation.
This mechanism allows us to adapt to future architectures as we can simply add new built-in instructions and their specifications to Fireiron, which is exactly what we did for both HMMA and WMMA.

\subsection{\emph{Cooperation}: Data Types for Parallel Cooperation. }
In parallel programs, multiple threads cooperate to (1) load arrays, (2) compute new values, and (3) store them back.
In advanced implementations, the mapping of the array elements to threads may change between these three phases.
To simplify programming such implementations, we want to abstract away from the mapping as long as possible, specifying it only at lower levels of the decomposition.
This makes higher-level specifications more general, permitting more possible implementations.
\paragraph{Mechanism:}
\emph{Distributed arrays} are data types that the programmer can distribute over private memories.
Fireiron provides the illusion that a distributed array is indivisibly stored across, say, the registers of multiple warps; here, registers are memories that are private to each thread.
The Fireiron compiler automatically divides the array and distributes the pieces onto registers of threads of the involved warps.
This is especially useful for the epilog of a kernel implementations where a CTA needs to move computed results (usually residing in registers spread across all its threads) back to global memory.

%% file: tex/specs.tex
\begin{figure*}
        \includegraphics[width=\linewidth]{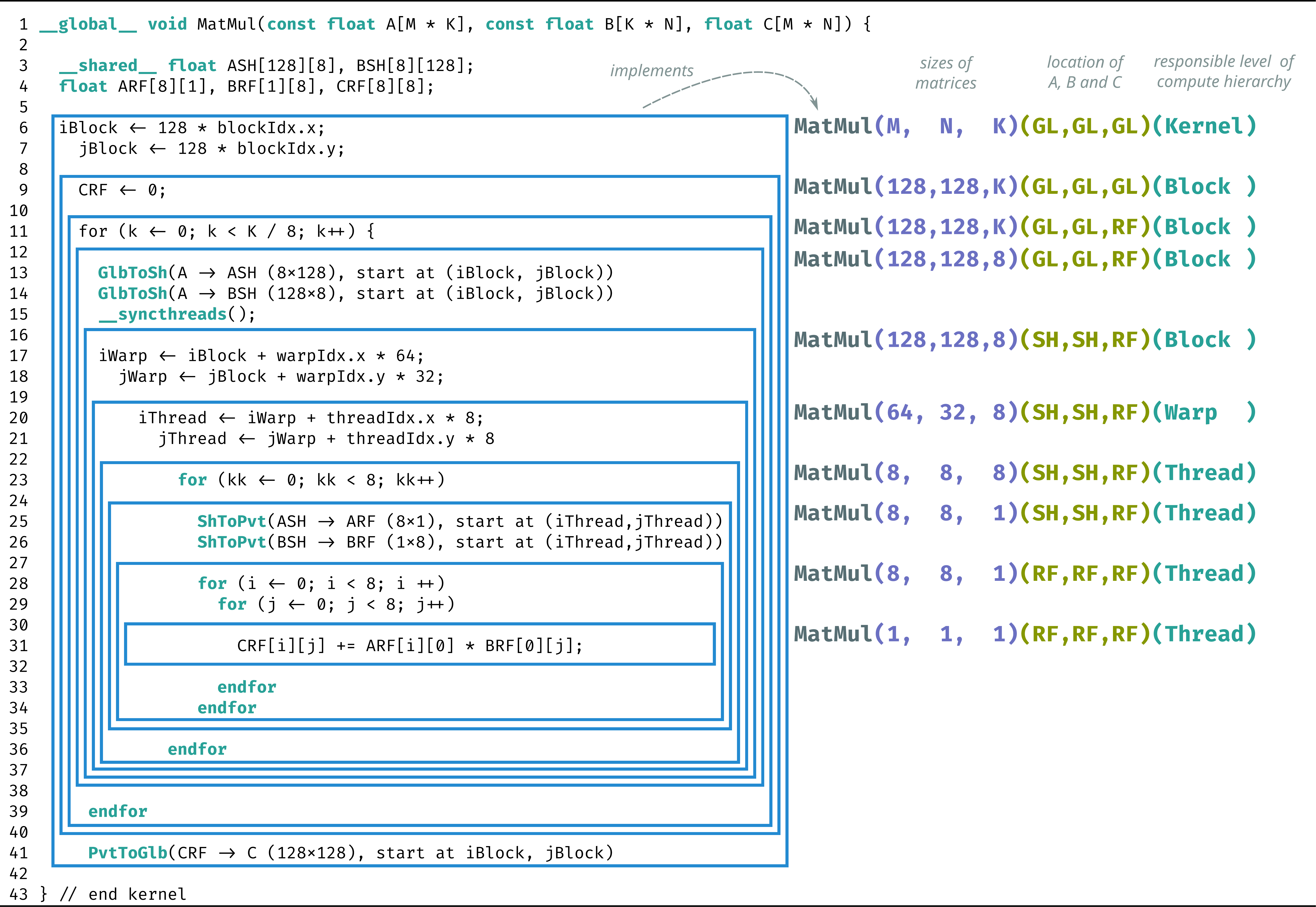}
        \caption{Visualization of the hierarchical structure of GPU programs using a matrix multiplication kernel as example. Within a typical GPU kernel we gradually descend the compute and memory hierarchy while computing smaller instances of the original problem.}
        \label{fig:boxes}
\end{figure*}
Most high-performance kernels for GPUs are written in a hierarchical style.
The original problem to be computed is decomposable into smaller sub-problems of the same kind.
These sub-problems are then assigned to and computed by the different levels of the compute hierarchy.
Figure~\ref{fig:boxes} visualizes this observation using a simple matrix multiplication kernel.
Step by step, the matrix multiplication is decomposed into a hierarchy of tiles and data is transferred to lower levels of the memory hierarchy until eventually every thread computes a single FMA instruction.
Here, FMA can be viewed as a matrix multiplication of matrices which only contain a single element.

Fireiron introduces two main concepts: \textit{Specifications} and \textit{Decompositions}.
These two concepts allow to describe implementations, such as the one shown in Figure~\ref{fig:boxes}, and their mapping to the hardware in a natural way.

\subsection{Specifications}
A \emph{Specification} (spec) is a data-structure describing the computation to implement.
A spec contains enough information such that a programmer would be able to manually write an implementation.
This especially entails that a spec keeps track of the shapes, locations and storage layouts of its input and output tensors, as well as which level of the compute hierarchy (i.e., Kernel, Block, Warp or Thread) is responsible for computing this operation.
Currently, Fireiron supports two main classes of specs: Matrix Multiplication (\texttt{MatMul}), and data movement (\texttt{Move}).
The following listing shows a kernel-level matrix multiplication spec:
\begin{lstlisting}[style=fireiron, language=Fireiron]
MatMul(ComputeHierarchy: Kernel,
       A: Matrix((M x K), float, GL, ColMajor),
       B: Matrix((K x N), float, GL, ColMajor),
       C: Matrix((M x N), float, GL, ColMajor))
\end{lstlisting}
At the beginning of every GPU kernel, inputs are stored in global memory (GL).
For matrices, Fireiron supports both static shapes (compile-time constants) and symbolic shapes denoted as simple arithmetic expressions e.g., \texttt{M = ((x + y) \% z)} where $x$, $y$ and $z$ are only known at runtime.
If not further specified, we assume that all matrices are stored in column-major format and contain elements of type float and write \texttt{MatMul(M,N,K)(GL,GL,GL)(Kernel)} as a short form representing the spec in the listing above.

Given a spec, one can perform one of the following actions:
1. Arrive at an executable spec; or
2. Decompose it into a smaller sub-spec.

\subsection{Executable Specifications}
A specification is called \emph{executable} when it matches the specification of a user-provided micro-kernel, or a built-in instruction.
Fireiron, contains a built-in collection of executable specs matching different instructions like FFMA and HFMA.
For example, the FMA instruction has the spec \texttt{MatMul(1,1,1)(RF,RF,RF)(Thread)}.
When a final implementation contains executable leaf specs, Fireiron will emit the matching built-in instruction or chosen micro-kernel when generating the implementation.

\paragraph{Micro-Kernels}
At any time when decomposing specs with Fireiron, the user can provide a handwritten micro-kernel which implements the current spec.
This allows the Fireiron user to break out of the DSL and use custom implementations, potentially written in low-level assembly, for which we cannot yet provide good abstractions.

\subsection{Decompositions}
\begin{figure}
        \includegraphics[width=\linewidth]{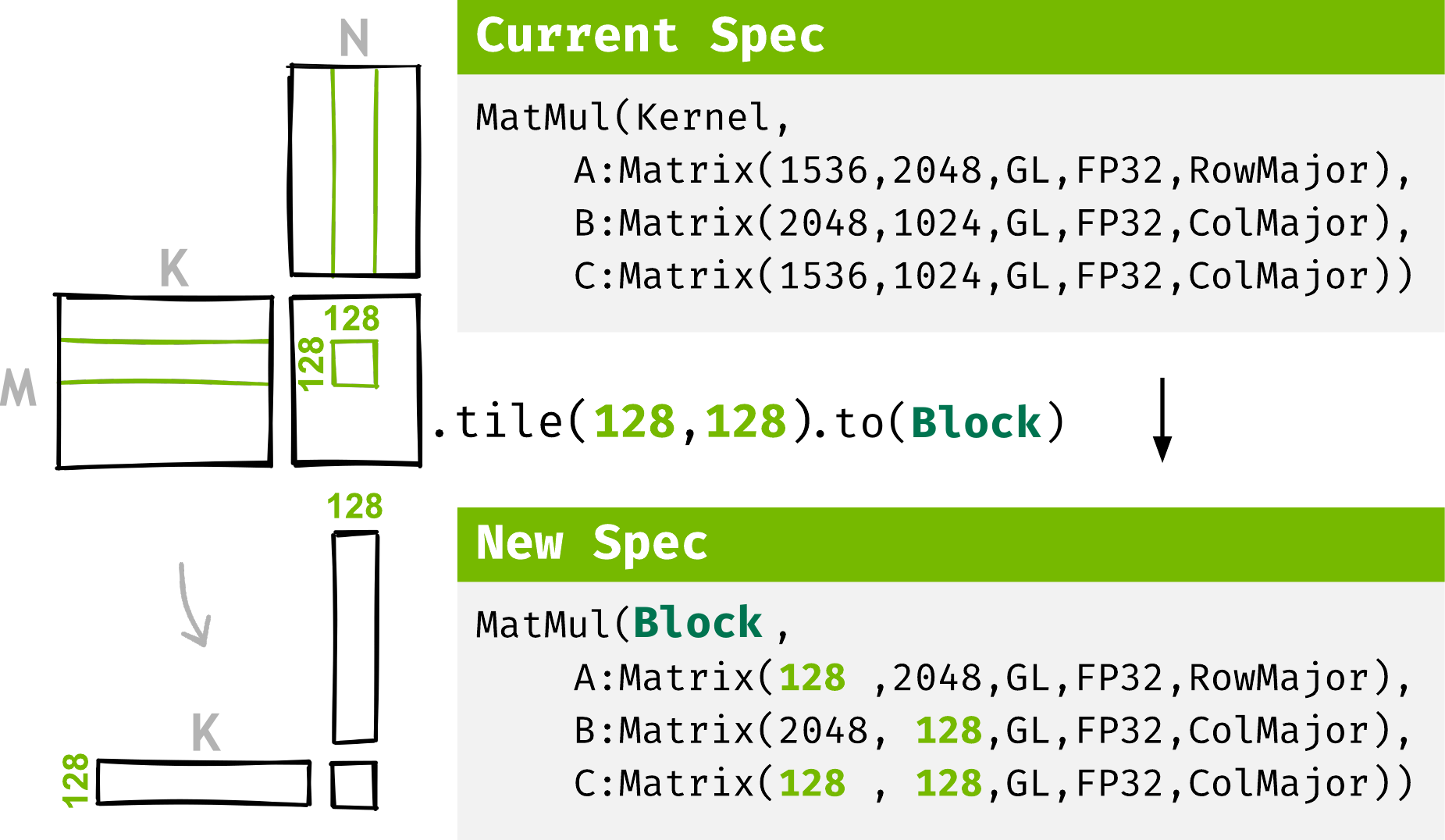}
        \caption{Tiling a \texttt{MatMul} spec results in a decomposed subspec with adjusted dimensions and optionally adjusted compute hierarchy to indicate parallel execution.}
        \label{fig:tile-viz}
\end{figure}
\begin{figure}
        \includegraphics[width=\linewidth]{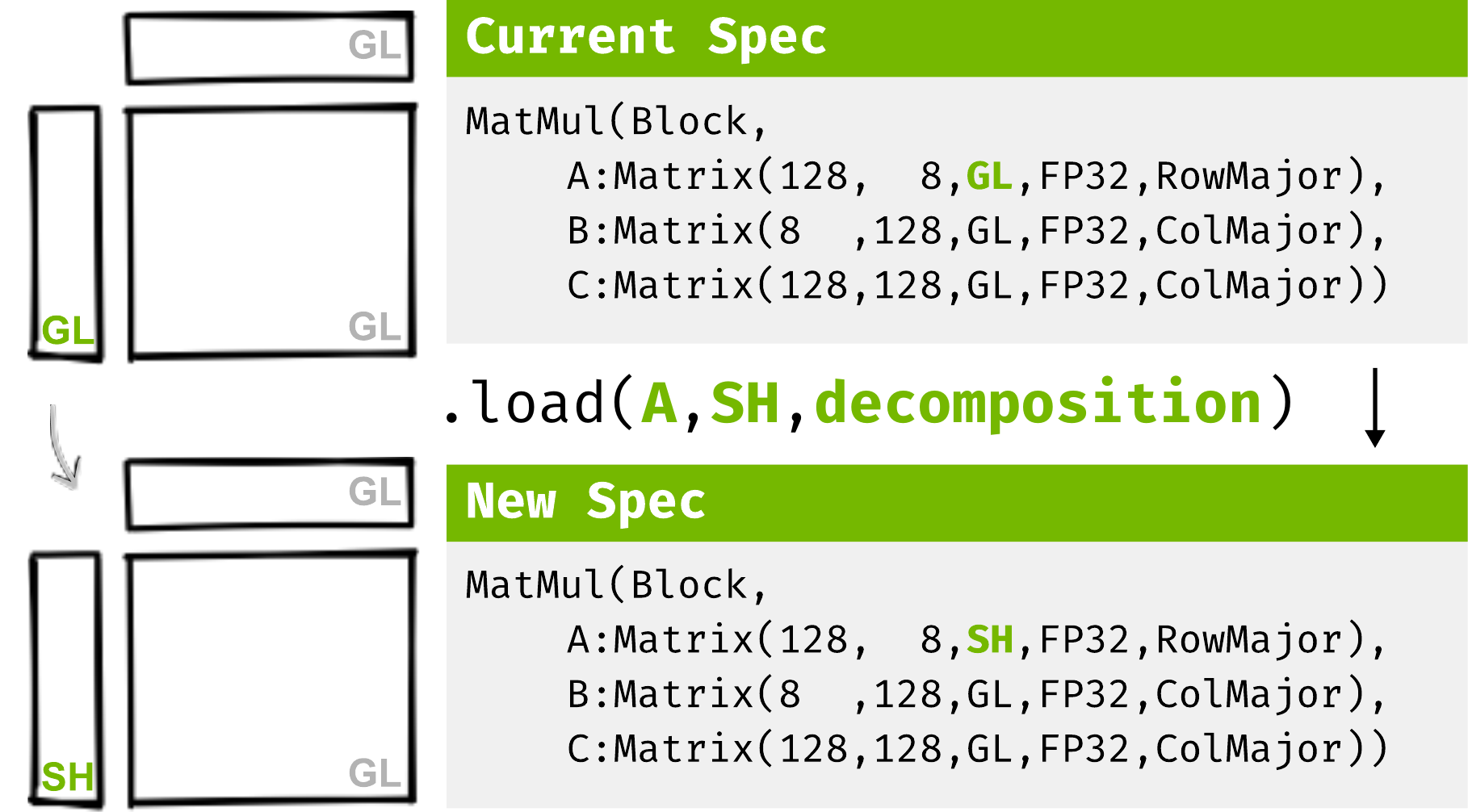}
        \caption{Applying \texttt{load} to a \texttt{MatMul} spec results in a new spec in which the memory location of the specified operand has changed. The load will be implemented as specified in the load-\emph{decomposition}.}
        \label{fig:load-viz}
\end{figure}

A \emph{Decomposition} describes how to (partially) implement a given spec.
More specifically, a decomposition is a function $\texttt{Spec} \rightarrow \texttt{Spec}$ which, given a spec, returns a new spec that represents the smaller decomposed sub-problem.
Fireiron provides two main decompositions, \texttt{tile} and \texttt{load}, which allow implementations to use the compute and memory hierarchy of a GPU.
\begin{itemize}
    \item \texttt{spec.tile(r,c)} creates $r \times c$ shaped tiles in the output matrix. Input matrices are tiled accordingly.
    In order to assign tiles to a level of the compute hierarchy, we can further \emph{refine} the tiling by applying \texttt{.to(level)} which changes the responsible compute hierarchy level for the resulting tiled spec.
    Figure~\ref{fig:tile-viz} shows the effect of tiling a \texttt{MatMul} spec.
    \item \texttt{spec.load(M,l,d)} loads the matrix $M$ to the level $l$ of the memory hierarchy with an implementation described as decomposition $d$.
    Figure~\ref{fig:load-viz} shows the effect of loading a \texttt{MatMul} spec.
    The memory hierarchy has three levels: global memory (GL), shared memory (SH) and registers (RF).
\end{itemize}
Fireiron is implemented as a domain-specific language, embedded in Scala, which generates CUDA.
We allow the user to define custom decompositions to allow them to implement advanced optimisations, as described later.

\subsection{Decomposing Matrix Multiplication}
Listing~\ref{lst:matmul101} shows an example decomposition of a \texttt{MatMul} spec using only \texttt{tile}, and \texttt{load}. The \texttt{done} operator marks the end of the decomposition and invokes the code generator.
This example can already be compiled into a simple, correct implementation.
Here, the load decompositions, which describe how to implement the data movement, are omitted for brevity (denoted \texttt{\_}) .


Note that unlike in the implementation shown in Figure~\ref{fig:boxes}, the K-dimension remains unchanged and the location of the C-matrix has not changed.
The residual spec (\texttt{MatMul(1,1,K)(RF,RF,GL)(Thread)}) is not executable, but instead describes a simple dot-product which is implemented in the micro-kernel \texttt{dot.cu}, and passed as an optional argument to \texttt{done}.

This simple decomposition already establishes major design decisions for implementing matrix multiplication:
it defines
1) how many elements each level of the compute hierarchy will compute and,
2) how many elements of each operand are stored at which layer of the memory hierarchy, (which determines the overall memory consumption).
Further decompositions will preserve this mapping, which emphasizes Fireiron's methodology for developing high-performance GPU kernels:
Fireiron allows programmers to focus on one thing at a time and then to gradually improve unspecified or sub-optimal parts of the decomposition.
\begin{lstlisting}[float, style=fireiron, language=Fireiron, label=lst:matmul101, caption=A simple decomposition for the \texttt{MatMul} spec: Each decomposition yields a smaller sub-specification which is further implemented by the subsequent decompositions.]
mm = MatMul(M,N,K)(GL,GL,GL)(Kernel)
mm              // resulting intermediate specs below
.tile(128,128)  // MatMul(128,128,K)(GL,GL,GL)(Kernel)
  .to(Block)    // MatMul(128,128,K)(GL,GL,GL)(Block )
.load(A, SH, _) // MatMul(128,128,K)(SH,GL,GL)(Block )
.load(A, SH, _) // MatMul(128,128,K)(SH,SH,GL)(Block )
.tile(64,32)    // MatMul(64, 32, K)(SH,SH,GL)(Block )
  .to(Warp)     // MatMul(64, 32, K)(SH,SH,GL)(Warp  )
.tile(8,8)      // MatMul(8,  8,  K)(SH,SH,GL)(Warp  )
  .to(Thread)   // MatMul(8,  8,  K)(SH,SH,GL)(Thread)
.load(A, RF, _) // MatMul(8,  8,  K)(RF,SH,GL)(Thread)
.load(B, RF, _) // MatMul(8,  8,  K)(RF,RF,GL)(Thread)
.tile(1,1)      // MatMul(1,  1,  K)(RF,RF,GL)(Thread)
.done(dot.cu)   // invoke codegen, emit dot micro-kernel
\end{lstlisting}

Every decomposition needs to specify three things which also need to be specified when adding new decompositions to Fireiron:
\begin{enumerate}
        \item[a.] Given the current spec, how to compute a new spec;
        \item[b.] The code to emit during code generation when processing this decomposition; and
        \item[c.] How to update the view into the current slices of the spec's operands to be able to generate correct indexing expressions.
\end{enumerate}

\paragraph{a. Computing the Sub-Spec}
Every application of a decomposition yields a new spec.
Listing~\ref{lst:matmul101} shows the intermediate specs after applying \texttt{tile} and \texttt{load}.
When applying \texttt{tile}, only the $M$ and $N$ dimension of all operands change.
Conversely, when applying \texttt{load}, only the memory location of the specified matrix changes.


\paragraph{b. Code Generation}
During code generation, we process the current decomposition tree from top to bottom, and insert code for every decomposition, until we reach the residual spec.
Generally, every decomposition represents one or more (potentially parallel) loop-nests whose body is defined by the subsequent decompositions.
For example in case of \texttt{tile}, two for-loops are generated which iterate over the created tiles:
\begin{lstlisting}[style=fireiron, language=C]
// .tile(mBlock, nBlock) emits:
for( int row = 0; row < mBlock; i++ ) {
  for( int col = 0; col < nBlock; j++ ) {
    // implementation of resulting spec
    subspec.codegen(); }}
\end{lstlisting}
If a level of the compute hierarchy has been assigned (using the \texttt{.to(level)} refinement), instead of emitting sequential for-loops, we use the unique compute hierarchy indexes for the specified level to assign a tile to each unit at that level:
\begin{lstlisting}[style=fireiron, language=C]
// .tile(mBlock, nBlock).to(Block) emits:
if(blockIdx.x < mBlock) {
  if( blockIdx.y < nBlock ) { subspec.codegen(); }}
\end{lstlisting}
For the \texttt{load} decomposition, we allocate a temporary array in the required memory hierarchy level at the beginning of the kernel and emit loops to copy the data to the new memory region:
\begin{lstlisting}[style=fireiron, language=C]
// .load(A, SH, d) emits:
for( /* iterate over A */ ) {
  // copy elements as specified in d
  A_SH[...] = A[...] // implementing: Move(A, GL->SH)
}
__syncthreads();   // if location == SH
subspec.codegen(); // implementation of resulting spec
\end{lstlisting}
Finally, if the residual spec is \emph{executable}, we emit the associated built-in instruction or micro-kernel (e.g. FMA or \texttt{dot.cu} in Listing~\ref{lst:matmul101}, respectively).

\paragraph{c. Index Computation}
Almost every optimization affects the computation of indexes in some way.
In Fireiron, index computations for accessing all matrices are calculated and emitted when required.
Every application of a decomposition returns a new spec in which we either sliced the operands or moved them to a new memory location.
In both cases, these changes need to be considered when generating index expressions for accessing the array elements.
For example, when applying \texttt{tile} to the \texttt{MatMul} spec, the following indices are computed as expected:
\begin{lstlisting}[style=fireiron, language=Fireiron]
// .tile(mBlock, nBlock) for MatMul:
C.rowIndex += rowVar * mBlock;
C.colIndex += colVar * nBlock;
A.rowIndex += rowVar * mBlock;
B.colIndex += colVar * nBlock;
\end{lstlisting}
When applying \texttt{load}, a fresh indexing expression for the newly allocated array is generated.

%% file: tex/optimization.tex
In this section, we show how to express advanced optimized strategies by refining the strategy shown in Listing~\ref{lst:matmul101}.

\subsection{Extending Specialized Decompositions}
Fireiron's existing decompositions can easily be extended by the user to express custom ways to decompose a given spec.
In this section, we introduce two matrix multiplication specific decompositions (\texttt{split} and \texttt{epilog}).
To add a new decomposition to Fireiron, we need to explain how to:
a) compute a new sub-spec;
b) generate a partial implementation;
c) update the index computations for the involved operands.

\paragraph{Splitting the K-Dimension}
For matrix multiplications, we need to be able to create tiles in the K-dimension. 
This allows the best use of shared memory, especially for large matrices, where a whole row of the A matrix (or column of B) might not fit into the limited shared memory.

Since the \texttt{split} decomposition is specific for matrix multiplication, it can only be applied to the \texttt{MatMul} spec, and its behavior is described as follows:
\begin{lstlisting}[style=fireiron, language=Fireiron]
mm1 = MatMul(M,N,K)(locA,locB,locC)(level)
mm2 = mm1.split(kBlock)
// mm2 == MatMul(M,N,kBlock)(locA,locB,locC)(level)
\end{lstlisting}
The following partial implementation will be emitted when processing \texttt{split} during code generation:
\begin{lstlisting}[style=fireiron, language=C]
// .split(kBlock) emits:
for(int k = 0; k < kBlock; k++) {
  subspec.codegen(); //implementation of sub-spec }
\end{lstlisting}
Finally, the index computations are updated as follows:
\begin{lstlisting}[style=fireiron, language=Fireiron]
A.colIndex += k * kBlock;
B.rowIndex += k * kBlock;
\end{lstlisting}

\paragraph{Specifying Efficient Epilogues}
The decomposition shown in Listing~\ref{lst:matmul101} did not change the memory location of C.
This is because without \texttt{split}, we write only once to C, namely after computing the dot-product of a whole row of A and column of B.
In an efficient implementation however, the K-dimension is split into chunks and outer-products are accumulated in registers.
Once we finish iterating over the chunks of the K-dimension, every CTA contains the final results of its assigned tile in registers spread across its threads.
In the simplest case, every thread stores its results to the required position in global memory.
However, depending on the memory layout of the operand matrices, it can be more efficient to cooperate with other threads of the same CTA to accumulate partial results in shared memory, before storing them to global memory.

Fireiron supports the concept of \emph{distributed arrays} and provides the illusion that we have an indivisible CTA-level matrix C, even though the actual matrix is distributed across registers private to each thread.
This allows separate decompositions for storing back the computed results from registers to global memory to implement, for example, the more efficient cooperation via shared memory.
Typical libraries cannot offer this kind of composability since we cannot pass this distributed CTA-level matrix to a procedure which implements the store because the location of the accumulation registers cannot be taken.

%

\paragraph{Epilog}
In order to express advanced decompositions for storing computed results, as well as accumulating intermediate results in registers, we introduce a new decomposition \texttt{.epilog(l,i,d)}.
Similar to \texttt{load}, \texttt{i} and \texttt{d} are decompositions.
Here, \texttt{i} describes the \emph{initialization} of a new buffer in \emph{location} \texttt{l} (usually in registers) used for accumulating the results.
The decomposition \texttt{d} describes the implementation of the \texttt{Move} spec which represents storing the results from \texttt{l} to C's original location (usually global memory).
The behavior of \texttt{epilog} is described as:
\begin{lstlisting}[style=fireiron, language=Fireiron]
mm1 = MatMul(M,N,K)(locA,locB,locC)(level)
mm2 = mm1.epilog(l,i,d)
// mm2 == MatMul(M,N,K)(locA,locB,l)(level)
\end{lstlisting}
Since Fireiron's decomposition language is hierarchical, the subsequent decompositions only need to know the new location of C.
Similar to computing index expressions for \texttt{load}, we start with a fresh index expression for accessing the newly allocated buffer.
The emitted code snippet for \texttt{epilog} is as follows:
\begin{lstlisting}[style=fireiron, language=C]
for( /* iterate over M,N */ ) { // init
  C_l[...] = 0; }  // initialize buffer in location l
subspec.codegen(); // impl (storing results in C_l)
for( /* iterate over M,N */ ) { // store
  C[...] = C_l[...]; } // copy elements as specified in d
\end{lstlisting}

\subsection{Advanced Optimization using Refinements}
\begin{lstlisting}[style=fireiron, language=Fireiron,float,caption=Fireiron Decomposition which represents the implementation shown in Figure~\ref{fig:boxes}, label=lst:sndDecomp]
mm = MatMul(M,N,K)(GL,GL,GL)(Kernel)
mm               // resulting intermediate specs below
.tile(128,128)   // MatMul(128,128,K)(GL,GL,GL)(Kernel)
  .to(Block)     // MatMul(128,128,K)(GL,GL,GL)(Block )
.epilog(RF,_,_)  // MatMul(128,128,8)(GL,GL,RF)(Block )
.split(8)        // MatMul(128,128,8)(GL,GL,RF)(Block )
.load(A, SH, _)  // MatMul(128,128,8)(SH,GL,RF)(Block )
.load(A, SH, _)  // MatMul(128,128,8)(SH,SH,RF)(Block )
.tile(64,32)     // MatMul(64, 32, 8)(SH,SH,RF)(Block )
  .to(Warp)      // MatMul(64, 32, 8)(SH,SH,RF)(Warp  )
.tile(8,8)       // MatMul(8,  8,  8)(SH,SH,RF)(Warp  )
  .to(Thread)    // MatMul(8,  8,  8)(SH,SH,RF)(Thread)
.split(1)        // MatMul(8,  8,  1)(SH,SH,RF)(Thread)
.load(A, RF, _)  // MatMul(8,  8,  1)(RF,SH,RF)(Thread)
.load(B, RF, _)  // MatMul(8,  8,  1)(RF,RF,RF)(Thread)
.tile(1,1)       // MatMul(1,  1,  1)(RF,RF,RF)(Thread)
.done            // codegen emits FMA for residual spec
\end{lstlisting}


Listing~\ref{lst:sndDecomp} shows how to use the decompositions to express the implementation from Figure~\ref{fig:boxes}.
A strategy like this already establishes the sizes of data assigned to all levels of the compute and memory hierarchy.
In order to generate high-performance kernels, however, we need to specialize all parts of the decomposition to optimally make use of the target hardware.

This is where the traditional approach of optimizing GPU kernels quickly becomes tedious.
Conceptually trivial changes, like changing storage layouts during loads or using inline PTX rather than CUDA, require disproportionate amounts of work since a large fraction of the kernel will need to be rewritten.
Fireiron provides easy-to-use refinements to liberate the programmer from these tedious tasks and allows them productively focus on achieving the best possible performance.

Refinements are optional modifications to decompositions which enable advanced optimizations.
For example, \texttt{to(level)} is a refinement we've already seen for the \texttt{tile} decomposition.
Without this refinement, we generate sequential for-loops; with it however, we assign a specific level of the compute hierarchy to the newly created tiles, effectively computing them in parallel.
In this fashion, Fireiron provides multiple easy-to-use refinements which allow kernels to be gradually fine-tuned to achieve optimal performance.

\paragraph{Tile Refinements}
The following refinements are available for tile decompositions:
\begin{itemize}
    \item \texttt{.to(level)} assigns the created tiles to the specified \texttt{level} of the compute hierarchy.
    \item \texttt{.unroll} unrolls the generated for-loops.
    \item \texttt{.layout(l)} assigns the tiles to elements of the compute hierarchy in either column- or row-major order.
            This allows the programmer to match the storage layout so array accesses are coalesced..
    \item \texttt{.swizzle(perm)} introduces the use of advanced swizzle expressions to further optimize the mapping of tiles to elements of the current level of the compute hierarchy.
\end{itemize}
The \texttt{layout} and \texttt{swizzle} refinments enable changing the mapping of tiles to blocks, warps and threads.
They allow the programmer to use different tile shapes for different operations at the same level of the compute hierarchy (such as independent loads of the A and B matrices).
This allows advanced optimisations where the assignment of data to threads is more complex than merely column- or row-major.

\paragraph{Load Refinements}
Fireiron contains the following refinements for load decompositions:
\begin{itemize}
    \item \texttt{.noSync} avoids emitting \texttt{\_\_syncthreads()} when loading to shared memory, to avoid potentially unnecessary synchronization.
    \item \texttt{.storageLayout(l)} stores the elements in the destination buffer using either row- or column-major storage layout.
    \item \texttt{.pad(n)} allocates $n$ extra columns in the destination buffer, to avoid memory bank conflicts.
    \item \texttt{.align(n)} ensures a specific alignment for the created destination buffer.
    \item \texttt{.reuseBuffer} reuses a previously allocated buffer in the same memory location if it is no longer used, to reduce memory usage.
\end{itemize}

\paragraph{Split Refinements}
Fireiron also allows the user to define refinements for custom decompositions.
This is done by registering the new refinements, including their required code snippets, in the code generator.
For the \texttt{split} decomposition, for example, we add two refinements:
\begin{itemize}
    \item \texttt{.unroll} unrolls the created for-loop, like the equivalent refinement on \texttt{tile}.
    \item \texttt{.sync} emits \texttt{\_\_syncthreads()} as the last statement in the body of the created for-loop. This may be required depending on how shared memory is used in a particular implementation.
\end{itemize}
We are aware that some refinements, especially \texttt{noSync} and \texttt{sync}, can allow incorrect implementations due to race conditions.
However, a decomposition without refinements will always generate correct code.
Until now these issues have not caused problems as Fireiron has only been used by performance experts.
However, we intend to improve the analyses within Fireiron to ensure these refinements cannot cause correctness issues.

\subsection{Instructions with varying Granularity}
In order to support tensor cores on newer GPUs, we have added new executable specs to Fireiron which represent the specialized MMA (Matrix Multiply Accumulate) operations.

\paragraph{Supporting WMMA}
\label{sec:wmma}
The new WMMA-API in CUDA 10.0 introduces warp-wide matrix multiply operations which operate on warp-wide register collections called \emph{fragments}.
In order to generate kernels which use the new WMMA API, we extend Fireiron in two ways:
First, we extend Fireiron's memory hierarchy and add a new level \texttt{Fragment<M,N,K>} (parameterized due to the CUDA API) in between shared memory and registers.
This allows data to be loaded into the fragments required by the new WMMA operation:
\begin{lstlisting}[style=fireiron, language=Fireiron]
MatMul(16,16,16)(SH,SH,SH)(Warp).load(A, FR<16,16,16>, d)
// == MatMul(16,16,16)(FR<16,16,16>,SH,SH)(Warp)
\end{lstlisting}
Second, we add three new built-in instructions to Fireiron:
\begin{lstlisting}[style=fireiron, language=Fireiron]
// wmma:mma_sync ==
MatMul(16,16,16)(FR,FR,FR)(Warp) // using FP16
// wmma::load_matrix_sync ==
Move(level: Warp,
     src: Matrix((M x N), FP16, _, _)
     dst: Matrix((M x N), FP16, FR _))
// wmma::store_matrix_sync ==
Move(level: Warp,
     src: Matrix((M x N), FP16, FR, _)
     dst: Matrix((M x N), FP16, _, _))
\end{lstlisting}
These small additions allow us to write a simple Fireiron decomposition which uses the WMMA API, as shown in Listing~\ref{lst:wmma-decomp-simple}.
Note that this code does not apply any further decompositions.
It computes the matrix multiplication as follows:
1) Assign $64\times64$ elements to a CTA (line 5); 2) Initialize 16 ($4\times4$) accumulator fragments (line 7); 3) fill operand fragments (lines 12-13); compute the result (line 14); and store results from fragments to global memory (line 8).
Note that we only need to decompose the computation until we reach the warp level because we can generate code for the residual executable spec (line 14) using the built-in \texttt{wmma::mma\_sync} instruction.

\begin{lstlisting}[style=fireiron, language=Fireiron, numbers=left, float, caption=Simple WMMA decomposition describing the implementation of the first cudaTensorCoreGemm kernel shown in the CUDA samples., label=lst:wmma-decomp-simple]
val fragment = FR<16,16,16>
///// MATMUL-KERNEL /////////////////////////////////
val simpleWMMA = MatMul
///// BLOCK-LEVEL ///////////////////////////////////
  .tile(64, 64).to(Block)
  .epilog(fragment, Float,
    Move.tile(16, 16).to(Warp).done, // init
    Move.tile(16, 16).to(Warp).done) // store
  .split(16)
///// WARP-LEVEL ////////////////////////////////////
  .tile(16, 16).to(Warp)
  .load(MatMul.a, fragment, Move.done)
  .load(MatMul.b, fragment, Move.done)
  .done // => MatMul(16,16,16)(FR,FR,FR)(Warp)
\end{lstlisting}

\paragraph{Supporting HMMA}
Using the HMMA instructions exposed via PTX~\footnote{https://docs.nvidia.com/cuda/parallel-thread-execution/index.html\#warp-level-matrix-instructions-mma} allows even more fine-grained control over how the tensor cores are used.
In order to support these instructions, we extend Fireiron with executable specs which exactly describe their semantics:
\begin{lstlisting}[style=fireiron, language=Fireiron]
// HMMA.884.F16.TN exectuable spec
MatMul(ComputeHierarchy: Thread,
       A: Matrix((1x4), FP16, RF, RowMajor),
       B: Matrix((4x1), FP16, RF, ColMajor),
       C: Matrix((1x8), FP16, RF, ColMajor))
\end{lstlisting}
Note the unusal shapes of the operands which are dictated by the semantics of the \texttt{HMMA} instruction.

None of the existing decompositions can be used to create such slices of operands.
In order decompose a spec into this executable spec, we need to add an additional mma-specific decomposition.
The \texttt{mmaTile} decomposition expects a contiguous 16$\times$16 warp-level MatMul spec and returns an executable HMMA spec.

%% file: tex/evaluation.tex
To evaluate Fireiron, we compare the performance of our generated code against three references:
1) a manually-tuned kernel targeting the Maxwell architecture written by NVIDIA's GPU performance experts,
2) the publicly available CUDA sample \emph{cudaTensorCoreGemm} kernel\footnote{https://github.com/NVIDIA/cuda-samples/blob/master/Samples/ cudaTensorCoreGemm/cudaTensorCoreGemm.cu} targeting the WMMA API, which exploits NVIDIA's TensorCores on Volta and Turing architectures, and
3) the high-performance GEMM implementations provided in cuBLAS which are written in low-level assembly and are the fastest GEMM implementations available for NVIDIA GPUs today.

We chose this set of comparisons to highlight Fireiron's capability to generate efficient code for different GPU architectures ranging from older, such as Maxwell, up to state-of-the-art, such as Turing.
Furthermore, this set of references includes a wide variety of optimizations which are all expressible using Fireiron.
These range from simple tiling and shared memory padding (to avoid bank conflicts), as used in the CUDA sample code, to carefully-tuned swizzles~\cite{DBLP:conf/asplos/PhothilimthanaE19} and inline PTX assembly which achieves the highest performance possible.

We expressed all reference algorithms using decompositions in Fireiron and compare the achieved performance of our generated code on different architectures.
Specifically, we used three different GPUs: GeForce GTX 750 Ti (Maxwell), Quadro GV100 (Volta) and GeForce RTX 2080 Ti (Turing), CUDA-10.0, Driver Version 425.00 and compiled all kernels using \texttt{-O3 --use\_fast\_math -arch=sm\_XX} where \texttt{XX} $=$ 52,70, and 75 for Maxwell, Volta, and Turing respectively.
We locked the clocks to fixed frequencies, report the minimum kernel runtime of 1000 runs using \textit{nvprof} and omit data transfer time because we are only interested in the quality of our generated kernel code.

\subsection{Targeting the Maxwell Architecture}
The first kernel we compare against is manually tuned for larger input sizes (M,N,K >= 1024) and optimized to run efficiently on the Maxwell architecture.
Listing~\ref{lst:maxwell-decomp} shows the decomposition used to express this specific implementation.
Note that we express two different load strategies for prefetching operands A (lines~\ref{line:maxwell-fetch-a-start}--\ref{line:maxwell-fetch-a-end}) and B (lines~\ref{line:maxwell-fetch-b-start}--\ref{line:maxwell-fetch-b-end}) to shared memory.
This is due to computing GEMM\_NT where one operand is transposed and in order to perform efficient loads, we need to consider the storage layout for both operands such that global memory loads are coalesced.
We use advanced swizzle expressions (line~\ref{line:maxwell-swizzle}) shuffling the mapping of data to threads to avoid shared memory load conflicts.
Furthermore, we make heavy use of refinements to explicitly specify which loops to unroll and where to add or avoid synchronization.
This decomposition also uses vectorized loads (lines~\ref{line:maxwell-vector-a} and~\ref{line:maxwell-vector-b}), sparse-thread tiles (line~\ref{line:maxwell-sparse}) and a custom epilog (line~\ref{line:maxwell-epilog}) which swizzles the data via shared memory, using an unconventional data-layout, to achieve efficient stores to global memory.

\begin{lstlisting}[float,style=fireiron, language=Fireiron, numbers=left,caption={Efficient decomposition for large input sizes on Maxwell. Optimizations expressed include: vectorized loads, sparse thread-tiles, swizzling, custom epilog micro-kernel.},label={lst:maxwell-decomp}]
val swizz: Swizzle = id => // permutation of thread-ids
  ((id >> 1) & 0x07) | (id & 0x30) | ((id & 0x01) << 3) @\label{line:maxwell-swizzle}@
// microkernel for epilog-store RF => GL
// implements spec: Move(128x128:RF => 128x128:GL)
val storeCUDA: String = //* CUDA code snippet *//
// MATMUL-KERNEL ////////////////////////////////////////
val maxwellOptimized = MatMul
///// BLOCK-LEVEL ///////////////////////////////////////
  .tile(128, 128).to(Block)
    .layout(ColMajor)
//--- epilog: store results RF => GL ------------------//
  .epilog(RF, @\label{line:maxwell-epilog}@
    Move // init 64 registers per thread
      .tile(64, 32).to(Warp)
      .tile(8, 8).to(Lane)
      .tile(1, 1).unroll
      .done,
    Move // store results using microkernel
      .done(storeCUDA))
  .split(8).sync
//--- load A to SH ------------------------------------//
  .load(MatMul.a, SH, Move @\label{line:maxwell-fetch-a-start}@
    .tile(128, 1).to(Warp)
    .tile(64, 1).unroll
    .tile(2, 1).to(Lane)
      .layout(ColMajor)
    .done).toColMajor.noSync @\label{line:maxwell-fetch-a-end}@
//--- load B to SH ------------------------------------//
  .load(MatMul.b, SH, Move @\label{line:maxwell-fetch-b-start}@
    .tile(8, 16).to(Warp)
    .tile(8, 4).unroll
    .tile(1, 1).to(Lane)
      .layout(ColMajor)
    .done).toRowMajor.pad(4) @\label{line:maxwell-fetch-b-end}@
///// WARP-LEVEL ////////////////////////////////////////
  .tile(64, 32).to(Warp)
///// THREAD-LEVEL //////////////////////////////////////
  .tile(Strided((4, 32), (4, 16))) @\label{line:maxwell-sparse}@
      .to(Lane)
      .layout(ColMajor)
      .swizzle(swizz)
  .split(1).unroll
//--- load A to RF ------------------------------------//
  .load(MatMul.a, RF,
    Move.tile(4, 1).unroll.done) @\label{line:maxwell-vector-a}@
//--- load B to RF ------------------------------------//
  .load(MatMul.b, RF,
    Move.tile(1, 4).unroll.done) @\label{line:maxwell-vector-b}@
//--- perform computation -----------------------------//
  .tile(1, 1).unroll
  .done // residual = MatMul(1,1,1)(RF,RF,RF)(Thread)
\end{lstlisting}

\begin{figure}
        \includegraphics[width=\linewidth]{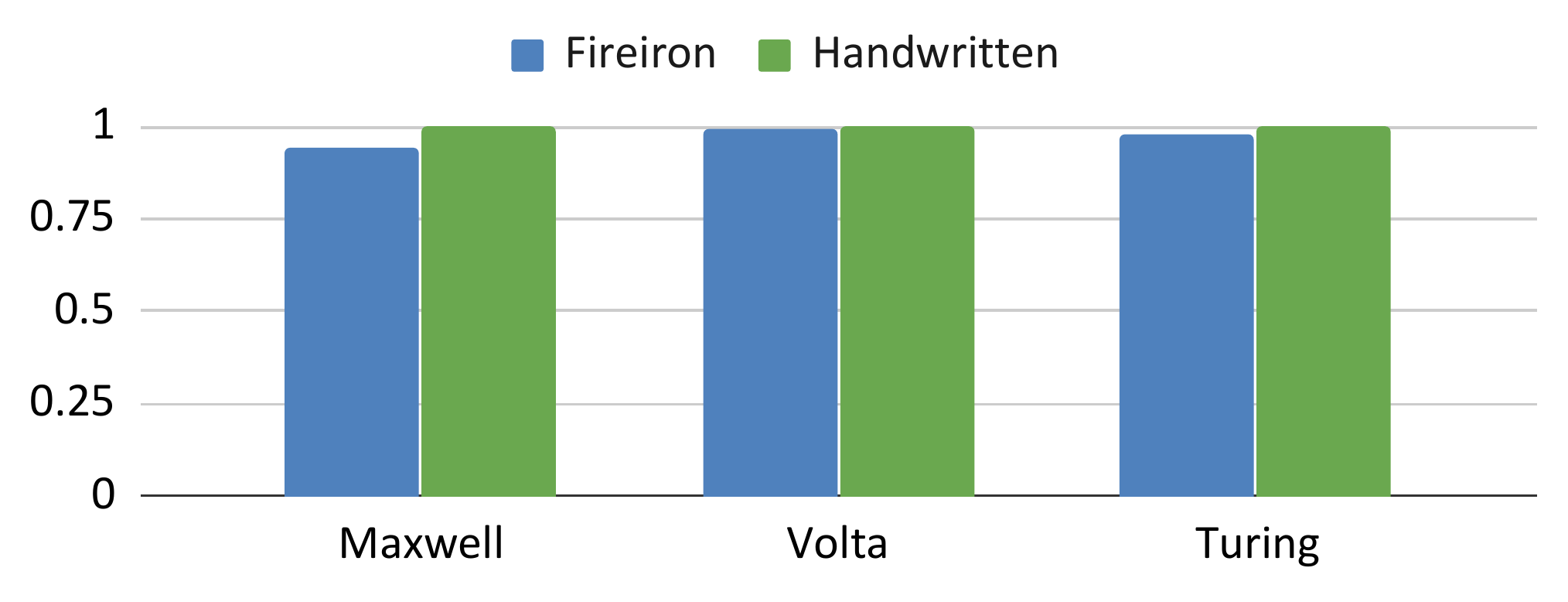}
        \caption{Comparing Fireiron generated kernel to manually tuned kernel targeting the Maxwell architecture. (computing HGEMM\_NT, M=1536, N=1024, K=2048)}
        \label{fig:maxwell}
\end{figure}
Figure~\ref{fig:maxwell} shows the achieved performance relative to the reference kernels (higher is better).
We executed both the reference and the Fireiron-generated kernels, on the Maxwell, Volta, and Turing architectures.
Our automatically generated code achieves 94.4\% of the available performance compared to the manually written CUDA code on Maxwell, as well as 98\% and 99\% on Volta and Turing respectively.
Due to automatic index expression generation, our loops and array indices look slightly different to the hand-written CUDA code we compared against, and can be further optimised, which explains the remaining gap in performance.

However, due to the tensor cores introduced with the Volta architecture, this implementation is now only of limited interest.
The Volta architecture supports the MMA instruction which has an 8$\times$ higher theoretical peak performance than the FMA instruction.
In fact, we are able to significantly outperform this specific implementation (up to 5$\times$) by avoiding the FMA instruction and using Fireiron decompositions better suited to the newer architectures.
Similarly, this GEMM implementation only achieves 37\% of the performance on Volta and 19\% of the performance on Turing when comparing to cuBLAS.
Still, this comparison highlights Fireiron's ability to express the exact same optimizations as manual implementations which nonetheless achieve the same performance.

\subsection{Targeting TensorCores using WMMA}
The NVIDIA CUDA samples contain two kernels which use WMMA to put the TensorCores to use.
The first kernel simply introduces the API calls and is not optimized.
In fact, the decomposition shown in Section~\ref{sec:wmma} generates exactly this kernel.
The second kernel is slightly optimized, using tiling and padded shared memory buffers to avoid bank conflicts.
Listing~\ref{lst:wmma-decomp} shows this implementation expressed as a Fireiron decomposition.
\begin{lstlisting}[float,style=fireiron, language=Fireiron,caption={Fireiron WMMA Decomposition. Note that we do not descend down to the thread-level as the WMMA-instruction is a warp-wide instruction.},label={lst:wmma-decomp}]
// In the following: FR == WMMA-Fragment
// MATMUL-KERNEL ////////////////////////////////////////
    val impl = MatMul
///// BLOCK-LEVEL ///////////////////////////////////////
      .tile(128, 128).to(Block)
//--- distributed store: FR => GL ---------------------//
      .epilog(FR,
        Move// init: WMMA-Fragment for C
          .tile(64, 32).to(Warp)
          .tile(16, 16).unroll.done,
        Move// store: FR => GL
          // FR => SH (first step)
          .load(Move.src, SH, Move
            .tile(64, 32).to(Warp)
            .tile(16, 16).unroll.done
          ).reuseBuffer
          // SH => GL (second step)
          .tile(16, 128).to(Warp)
          .tile(1, 128).unroll
          .tile(1, 4).to(Lane)
          .done)
      .split(128).sync.unroll
//--- load A to SH ------------------------------------//
      .load(MatMul.a, SH, Move
        .tile(16, 128).to(Warp)
        .tile(2, 128).unroll
        .tile(1, 8).to(Lane)
        .done).noSync.pad(8)
//--- load B to SH ------------------------------------//
      .load(MatMul.b, SH, Move
        .tile(128, 16).to(Warp)
        .tile(128, 2).unroll
        .tile(8, 1).to(Lane).layout(ColMajor)
        .done).pad(8)
///// WARP-LEVEL ////////////////////////////////////////
      .tile(64, 32).to(Warp)
      .split(16).unroll
//--- fill WMMA fragments for A -----------------------//
      .load(MatMul.a, FR, Move
        .tile(16, 16).unroll.done)
//--- fill WMMA fragments for B -----------------------//
      .load(MatMul.b, FR, Move
        .tile(16, 16).unroll.done)
//--- perform WMMA computation ------------------------//
      .tile(16, 16).unroll
      .done // MatMul(16,16,16)(FR,FR,FR)(Warp)
\end{lstlisting}

\begin{figure}
    \includegraphics[width=\linewidth]{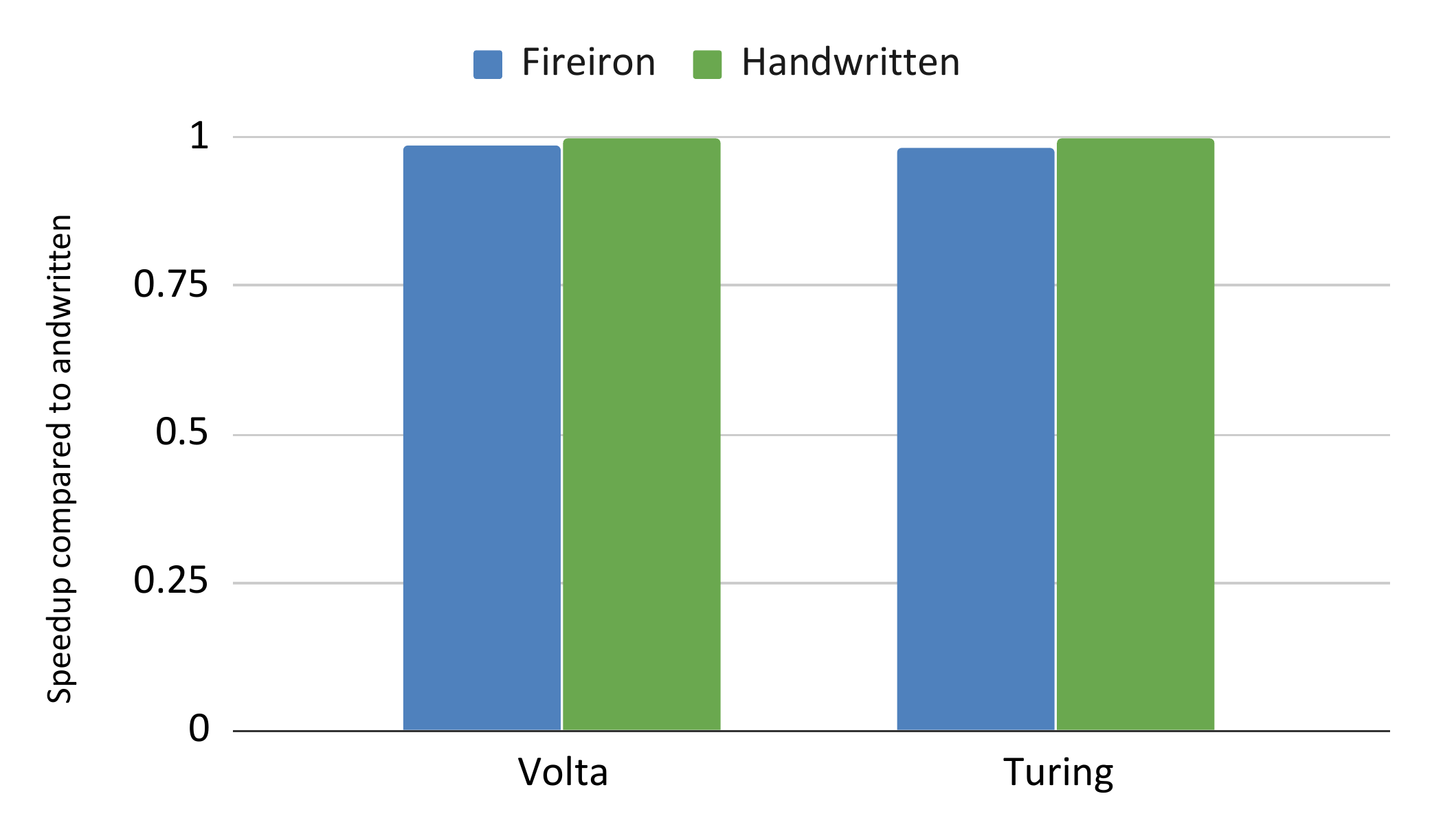}
        \caption{Comparing Fireiron generated Kernel to the optimized kernel in the CUDA SDK Samples (cudaTensorCoreGemm)}
        \label{fig:wmma-perf-volta}
\end{figure}

Figure~\ref{fig:wmma-perf-volta} shows the achieved performance for the optimized decomposition on both Volta and Turing.
Again, our decomposition expresses exactly the optimizations implemented by hand in the reference kernel.
Therefore, the generated kernel achieves the same performance as the manually written kernel, but the decomposition is much more concise than the optimized CUDA code.
In order to develop this decomposition, we able to start with the unoptimized decomposition and gradually refine it, focusing on adding one optimization at a time.

Compared to cuBLAS this implementation however still only achieves 61\% of the performance on Volta and 68\% of the performance on Turing.
This is because the memory and compute hierarchy need to be used more efficiently.
In particular, using PTX's \texttt{mma.sync} instruction, instead of the more coarse grained WMMA API, enables delicate control over the TensorCores computation.

%
\subsection{High-Performance HMMA Inline PTX}
\begin{figure}
    \includegraphics[width=\linewidth]{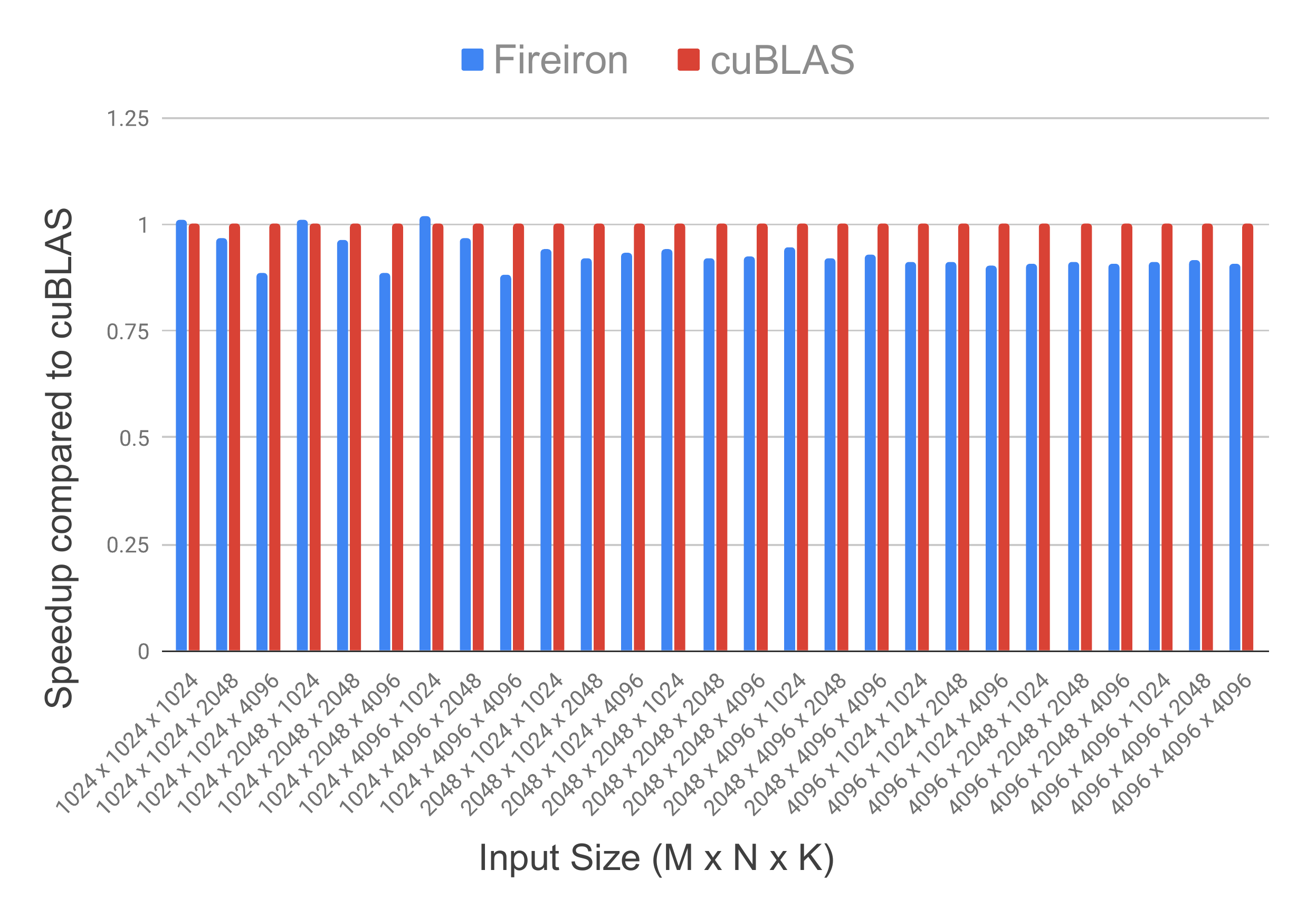}
        \caption{Relative Performance of Fireiron's inline PTX Decomposition compared to cuBLAS for large input matrices.}
        \label{fig:large-matrices}
\end{figure}
\begin{figure*}
    \includegraphics[width=\linewidth]{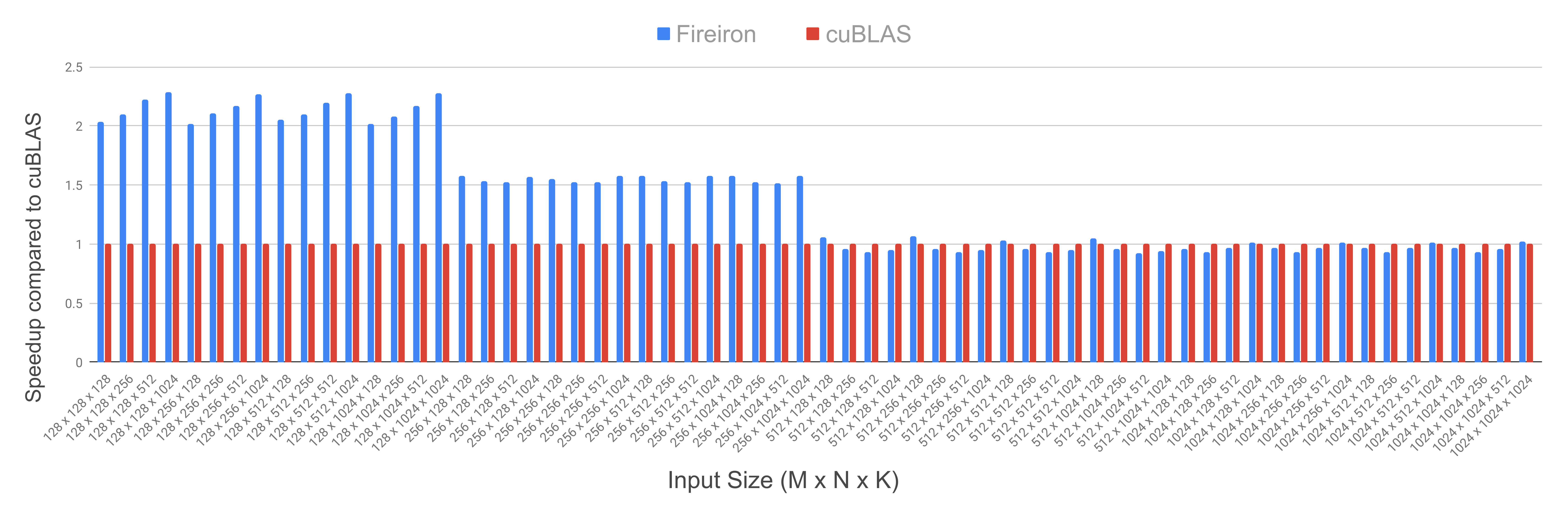}
        \caption{Relative Performance of Fireiron's inline PTX Decomposition compared to cuBLAS for small input matrices}
        \label{fig:small-matrices}
\end{figure*}
Finally, we focus on achieving the highest performance possible and compare against cuBLAS, which provides the most efficient GEMM implementations for several architectures and input sizes (written in optimized SASS assembly).
We compare against the \texttt{cublasGemmEx} routine using half precision floating-point, so the \texttt{mma} instruction is used on architectures with TensorCores.

We express the implementations by composing Fireiron's decompositions to use the \texttt{mmaTile} decomposition to target the executable spec for HMMA.
cuBLAS uses different CTA-tile sizes depending on the input size.
We parameterized our decompositions which allows us to choose different sizes.
Since we are interested in seeing whether Fireiron can express all optimizations required to achieve state-of-the-art performance on new architectures, we explore several input sizes in this experiment.
Specifically, we run two separate experiments where we explore small ($M,N,K \le 1024$) and large matrices ($1024 \le M,N,K  \le 4096$).
The optimizations required to achieve high performance on small and large input sizes differ significantly, as the small input sizes expose less parallelism opportunities.
cuBLAS provides multiple optimized implementations and chooses a specific one based on internal heuristics.
In order to have a fair comparison, instead of only generating one Fireiron kernel, we wrote two parameterized decompositions (one generally more suited for smaller and one for larger input sizes), exhaustively explored CTA-tile sizes (powers of two: $2^4$--$2^8$), and report the best performance.
Figure~\ref{fig:large-matrices} shows the achieved performance compared to cuBLAS for the large input matrices (higher is better), and Figure~\ref{fig:small-matrices} shows the achieved performance for small matrices.

Generally, we are able to significantly increase the performance compared to the previous Maxwell- and WMMA-implementations.
For large matrices, we exactly match the performance of the carefully tuned SASS kernels used in cuBLAS in three cases.
On average, we achieve 93.1\% of the cuBLAS performance with minimum of 88.3\% in one case and a maximum of 101\% in two cases.
These results show that Fireiron can produce state-of-the-art CUDA kernels which achieve performance very close to the practical peak performance.
This is emphasized by the fact that cuBLAS kernels are provided as optimized SASS assembly, which exposes further optimization opportunities unavailable to Fireiron's generated CUDA kernels.
We left out the decompositions for this experiment for brevity, but none made use of any micro-kernels.
Instead, we used \texttt{mmaTile} and the library of executable HMMA specs to inject inline PTX for appropriate residual specs.

For small input sizes ($< 1024$) cuBLAS typically uses implementations where the reduction of the K-dimension is additionally parallelized across warps (usually referred to as in-CTA-split-k).
This creates a three dimensional warp arrangement and exposes more parallelism in the cases where two dimensional tiles are not enough to saturate all cores.

The speedups we achieve for the smallest input sizes are due to the tile sizes we found to work best for this particular setup.
We generally found a tile size $16\times16$ in the M and N dimensions and 64 in the K dimension, computed by two warps per CTA, to perform best.
The heuristics of cuBLAS also chose 64 for the K dimension, but larger sizes for the M and N dimensions, which reduces the available parallelism required to achieve even better performance.

Developing high-performance kernels in Fireiron allows comfortable experimentation with optimizations.
With Fireiron, developers can solely focus on expressing optimizations while the compiler automates tedious tasks like computing array indices which are easy to get wrong and often need to change throughout the whole kernel.
This significantly increases programmer productivity and leads to high-quality GPU code.

%% file: tex/related.tex
\paragraph{Schedule-based Compilers}
Fireiron is heavily inspired by Halide~\cite{DBLP:journals/tog/Ragan-KelleyAPLAD12,DBLP:conf/pldi/Ragan-KelleyBAPDA13} and TVM~\cite{DBLP:conf/osdi/ChenMJZYSCWHCGK18} in using decompostitions (i.e., scheduling primitives) to express optimizations for specs.
Similar frameworks include Tiramisu~\cite{DBLP:conf/cgo/BaghdadiRRSAZSK19}, CHiLL~\cite{chen2008chill}, and for graph algorithms GraphIt~\cite{DBLP:journals/pacmpl/ZhangYBKSA18}.
Some frameworks also use C's pragmas and other annotations to create DSLs for manual loop optimisations~\cite{DBLP:conf/lcpc/DonadioBRYBCGPP05,DBLP:conf/iwomp/KruseF18,DBLP:conf/ipps/HartonoNS09}.
None of these, however, provides neither the same flexibility nor do they allow to decompose existing efficient library routines into reusable components.

Strategy languages~\cite{DBLP:conf/icfp/VisserBT98, DBLP:journals/ijfcs/BorovanskyKKR01, DBLP:conf/cc/BrandDHJJKKMOSVVV01} which orchestrate the application of rewrite rules in term rewriting systems can be seen as predecessors to today's schedule-based compilers.
However, to the best of our knowledge, these languages have not been used in the context of generating efficient GPU code which implements state-of-the-art optimizations and which can compete with the performance of vendor-tuned libraries.
Similar analogies can also be drawn between Fireiron and the use of tactics in theorem provers to manually decompose proofs into sub-goals~\cite{DBLP:conf/lpar/Delahaye00}, though our schedule DSL is not turing-complete and does not allow the same levels of reflection found in these tactic languages.

\paragraph{High-Performance Code Generation}
Currently, Fireiron provides optimized implementations for a set of specs.
Lift~\cite{DBLP:conf/cgo/HagedornSSGD18,DBLP:conf/cgo/SteuwerRD17}, TensorComprehension~\cite{DBLP:journals/corr/abs-1802-04730} and Futhark~\cite{DBLP:conf/pldi/HenriksenSEHO17} are frameworks also aiming at high-performance code generation.
In contrast to fixed specifications encoding the computations as used in Fireiron, these frameworks allow to specify computations using a flexible high-level (functional) programming language.
Rather than simply extending the set of available Fireiron specs we can imagine expressing computations in a similar pattern-based high-level programming language in the future.

\paragraph{Polyhedral Compilers}
Diesel~\cite{DBLP:conf/pldi/ElangoRRSG18}, NOVA~\cite{DBLP:conf/pldi/ElangoRRSG18} and PPCG~\cite{DBLP:journals/taco/VerdoolaegeJCGTC13} are compilers making heavy use of optimization via the polyhedral model.
Since most of our decompositions eventually result in nested loops with affine accesses, Fireiron could potentially profit from using polyhedral techniques too, especially as the resulting implementation has a representation similar to Schedule Trees~\cite{Verdoolaege2014impact}.
In contrast to polyhedral optimization techniques however, Fireiron's decomposition operate on a higher-level of abstraction.
Fireiron's decompositions directly modify specifications which later compile to low-level loops implemented in CUDA.

\paragraph{Auto-Tuning and Program Synthesis}
Auto-Tuning approaches including Halide's auto-tuners~\cite{DBLP:journals/tog/AdamsMABLGSJFDR19,DBLP:conf/pldi/Ragan-KelleyBAPDA13}, OpenTuner~\cite{DBLP:conf/IEEEpact/AnselKVRBOA14}, ATF~\cite{DBLP:conf/hpcc/RaschHG17} and MD-Hom~\cite{DBLP:conf/iwocl/RaschSG19}, and program synthesis techniques such as SwizzleInventor~\cite{DBLP:conf/asplos/PhothilimthanaE19} aim to automatically develop optimized implementations by navigating a search space of possible implementations.
We see potential for a similar automatic search space exploration for Fireiron`s decompositions, however as of today, Fireiron is designed as a tool for performance experts, simplifying the development of optimizations rather than automatically searching for highly optimized implementations.

%% file: tex/conclusion.tex
In this paper we introduced Fireiron, a scheduling language for high-performance linear algebra on GPUs. 
We introduced the concept of specifications, which represent a computation to optimize, and decompositions which partially implement them.
Defining low-level PTX assembly as well as marco-instructions like WMMA as executable specs allows us to flexibly target new architectures including TensorCores.

Using matrix multiplication as a case study, we showed how to develop high-performance implementations using Fireiron's specs, decompositions and refinements.
Fireiron is expressive enough to support all optimizations typically used in hand-tuned kernels and flexible enough to allow the insertion of micro-kernels when no suitable high-level abstractions can be built easily.
Our experimental evaluation shows that Fireiron generates code with performance competitive to vendor-tuned high-performance libraries.
Finally, all Fireiron programs are composed of building blocks which can be reused in future implementations targeting new hardware architectures, allowing these optimisations to be applied more widely.

%% file: main.bbl

\begin{thebibliography}{25}


\ifx \showCODEN    \undefined \def \showCODEN     #1{\unskip}     \fi
\ifx \showDOI      \undefined \def \showDOI       #1{#1}\fi
\ifx \showISBNx    \undefined \def \showISBNx     #1{\unskip}     \fi
\ifx \showISBNxiii \undefined \def \showISBNxiii  #1{\unskip}     \fi
\ifx \showISSN     \undefined \def \showISSN      #1{\unskip}     \fi
\ifx \showLCCN     \undefined \def \showLCCN      #1{\unskip}     \fi
\ifx \shownote     \undefined \def \shownote      #1{#1}          \fi
\ifx \showarticletitle \undefined \def \showarticletitle #1{#1}   \fi
\ifx \showURL      \undefined \def \showURL       {\relax}        \fi
\providecommand\bibfield[2]{#2}
\providecommand\bibinfo[2]{#2}
\providecommand\natexlab[1]{#1}
\providecommand\showeprint[2][]{arXiv:#2}

\bibitem[\protect\citeauthoryear{Adams, Ma, Anderson, Baghdadi, Li, Gharbi,
  Steiner, Johnson, Fatahalian, Durand, and Ragan{-}Kelley}{Adams
  et~al\mbox{.}}{2019}]%
        {DBLP:journals/tog/AdamsMABLGSJFDR19}
\bibfield{author}{\bibinfo{person}{Andrew Adams}, \bibinfo{person}{Karima Ma},
  \bibinfo{person}{Luke Anderson}, \bibinfo{person}{Riyadh Baghdadi},
  \bibinfo{person}{Tzu{-}Mao Li}, \bibinfo{person}{Micha{\"{e}}l Gharbi},
  \bibinfo{person}{Benoit Steiner}, \bibinfo{person}{Steven Johnson},
  \bibinfo{person}{Kayvon Fatahalian}, \bibinfo{person}{Fr{\'{e}}do Durand},
  {and} \bibinfo{person}{Jonathan Ragan{-}Kelley}.}
  \bibinfo{year}{2019}\natexlab{}.
\newblock \showarticletitle{Learning to optimize halide with tree search and
  random programs}.
\newblock \bibinfo{journal}{\emph{{ACM} Trans. Graph.}} \bibinfo{volume}{38},
  \bibinfo{number}{4} (\bibinfo{year}{2019}), \bibinfo{pages}{121:1--121:12}.
\newblock
\urldef\tempurl%
\url{https://doi.org/10.1145/3306346.3322967}
\showDOI{\tempurl}


\bibitem[\protect\citeauthoryear{Ansel, Kamil, Veeramachaneni, Ragan{-}Kelley,
  Bosboom, O'Reilly, and Amarasinghe}{Ansel et~al\mbox{.}}{2014}]%
        {DBLP:conf/IEEEpact/AnselKVRBOA14}
\bibfield{author}{\bibinfo{person}{Jason Ansel}, \bibinfo{person}{Shoaib
  Kamil}, \bibinfo{person}{Kalyan Veeramachaneni}, \bibinfo{person}{Jonathan
  Ragan{-}Kelley}, \bibinfo{person}{Jeffrey Bosboom},
  \bibinfo{person}{Una{-}May O'Reilly}, {and} \bibinfo{person}{Saman~P.
  Amarasinghe}.} \bibinfo{year}{2014}\natexlab{}.
\newblock \showarticletitle{OpenTuner: an extensible framework for program
  autotuning}. In \bibinfo{booktitle}{\emph{International Conference on
  Parallel Architectures and Compilation, {PACT} '14, Edmonton, AB, Canada,
  August 24-27, 2014}}. \bibinfo{pages}{303--316}.
\newblock
\urldef\tempurl%
\url{https://doi.org/10.1145/2628071.2628092}
\showDOI{\tempurl}


\bibitem[\protect\citeauthoryear{Baghdadi, Ray, Romdhane, Sozzo, Akkas, Zhang,
  Suriana, Kamil, and Amarasinghe}{Baghdadi et~al\mbox{.}}{2019}]%
        {DBLP:conf/cgo/BaghdadiRRSAZSK19}
\bibfield{author}{\bibinfo{person}{Riyadh Baghdadi}, \bibinfo{person}{Jessica
  Ray}, \bibinfo{person}{Malek~Ben Romdhane}, \bibinfo{person}{Emanuele~Del
  Sozzo}, \bibinfo{person}{Abdurrahman Akkas}, \bibinfo{person}{Yunming Zhang},
  \bibinfo{person}{Patricia Suriana}, \bibinfo{person}{Shoaib Kamil}, {and}
  \bibinfo{person}{Saman~P. Amarasinghe}.} \bibinfo{year}{2019}\natexlab{}.
\newblock \showarticletitle{Tiramisu: {A} Polyhedral Compiler for Expressing
  Fast and Portable Code}. In \bibinfo{booktitle}{\emph{{IEEE/ACM}
  International Symposium on Code Generation and Optimization, {CGO} 2019,
  Washington, DC, USA, February 16-20, 2019}}. \bibinfo{pages}{193--205}.
\newblock
\urldef\tempurl%
\url{https://doi.org/10.1109/CGO.2019.8661197}
\showDOI{\tempurl}


\bibitem[\protect\citeauthoryear{Borovansk{\'{y}}, Kirchner, Kirchner, and
  Ringeissen}{Borovansk{\'{y}} et~al\mbox{.}}{2001}]%
        {DBLP:journals/ijfcs/BorovanskyKKR01}
\bibfield{author}{\bibinfo{person}{Peter Borovansk{\'{y}}},
  \bibinfo{person}{Claude Kirchner}, \bibinfo{person}{H{\'{e}}l{\`{e}}ne
  Kirchner}, {and} \bibinfo{person}{Christophe Ringeissen}.}
  \bibinfo{year}{2001}\natexlab{}.
\newblock \showarticletitle{Rewriting with Strategies in {ELAN:} {A} Functional
  Semantics}.
\newblock \bibinfo{journal}{\emph{Int. J. Found. Comput. Sci.}}
  \bibinfo{volume}{12}, \bibinfo{number}{1} (\bibinfo{year}{2001}),
  \bibinfo{pages}{69--95}.
\newblock
\urldef\tempurl%
\url{https://doi.org/10.1142/S0129054101000412}
\showDOI{\tempurl}


\bibitem[\protect\citeauthoryear{Chen, Chame, and Hall}{Chen
  et~al\mbox{.}}{2008}]%
        {chen2008chill}
\bibfield{author}{\bibinfo{person}{Chun Chen}, \bibinfo{person}{Jacqueline
  Chame}, {and} \bibinfo{person}{Mary Hall}.} \bibinfo{year}{2008}\natexlab{}.
\newblock \bibinfo{booktitle}{\emph{CHiLL: A framework for composing high-level
  loop transformations}}.
\newblock \bibinfo{type}{{T}echnical {R}eport}.
  \bibinfo{institution}{Citeseer}.
\newblock


\bibitem[\protect\citeauthoryear{Chen, Moreau, Jiang, Zheng, Yan, Shen, Cowan,
  Wang, Hu, Ceze, Guestrin, and Krishnamurthy}{Chen et~al\mbox{.}}{2018}]%
        {DBLP:conf/osdi/ChenMJZYSCWHCGK18}
\bibfield{author}{\bibinfo{person}{Tianqi Chen}, \bibinfo{person}{Thierry
  Moreau}, \bibinfo{person}{Ziheng Jiang}, \bibinfo{person}{Lianmin Zheng},
  \bibinfo{person}{Eddie~Q. Yan}, \bibinfo{person}{Haichen Shen},
  \bibinfo{person}{Meghan Cowan}, \bibinfo{person}{Leyuan Wang},
  \bibinfo{person}{Yuwei Hu}, \bibinfo{person}{Luis Ceze},
  \bibinfo{person}{Carlos Guestrin}, {and} \bibinfo{person}{Arvind
  Krishnamurthy}.} \bibinfo{year}{2018}\natexlab{}.
\newblock \showarticletitle{{TVM:} An Automated End-to-End Optimizing Compiler
  for Deep Learning}. In \bibinfo{booktitle}{\emph{13th {USENIX} Symposium on
  Operating Systems Design and Implementation, {OSDI} 2018, Carlsbad, CA, USA,
  October 8-10, 2018.}} \bibinfo{pages}{578--594}.
\newblock
\urldef\tempurl%
\url{https://www.usenix.org/conference/osdi18/presentation/chen}
\showURL{%
\tempurl}


\bibitem[\protect\citeauthoryear{Delahaye}{Delahaye}{2000}]%
        {DBLP:conf/lpar/Delahaye00}
\bibfield{author}{\bibinfo{person}{David Delahaye}.}
  \bibinfo{year}{2000}\natexlab{}.
\newblock \showarticletitle{A Tactic Language for the System Coq}. In
  \bibinfo{booktitle}{\emph{Logic for Programming and Automated Reasoning, 7th
  International Conference, {LPAR} 2000, Reunion Island, France, November
  11-12, 2000, Proceedings}}. \bibinfo{pages}{85--95}.
\newblock
\urldef\tempurl%
\url{https://doi.org/10.1007/3-540-44404-1\_7}
\showDOI{\tempurl}


\bibitem[\protect\citeauthoryear{Donadio, Brodman, Roeder, Yotov, Barthou,
  Cohen, Garzar{\'{a}}n, Padua, and Pingali}{Donadio et~al\mbox{.}}{2005}]%
        {DBLP:conf/lcpc/DonadioBRYBCGPP05}
\bibfield{author}{\bibinfo{person}{S{\'{e}}bastien Donadio},
  \bibinfo{person}{James~C. Brodman}, \bibinfo{person}{Thomas Roeder},
  \bibinfo{person}{Kamen Yotov}, \bibinfo{person}{Denis Barthou},
  \bibinfo{person}{Albert Cohen}, \bibinfo{person}{Mar{\'{\i}}a~Jes{\'{u}}s
  Garzar{\'{a}}n}, \bibinfo{person}{David~A. Padua}, {and}
  \bibinfo{person}{Keshav Pingali}.} \bibinfo{year}{2005}\natexlab{}.
\newblock \showarticletitle{A Language for the Compact Representation of
  Multiple Program Versions}. In \bibinfo{booktitle}{\emph{Languages and
  Compilers for Parallel Computing, 18th International Workshop, {LCPC} 2005,
  Hawthorne, NY, USA, October 20-22, 2005, Revised Selected Papers}}.
  \bibinfo{pages}{136--151}.
\newblock
\urldef\tempurl%
\url{https://doi.org/10.1007/978-3-540-69330-7\_10}
\showDOI{\tempurl}


\bibitem[\protect\citeauthoryear{Elango, Rubin, Ravishankar, Sandanagobalane,
  and Grover}{Elango et~al\mbox{.}}{2018}]%
        {DBLP:conf/pldi/ElangoRRSG18}
\bibfield{author}{\bibinfo{person}{Venmugil Elango}, \bibinfo{person}{Norm
  Rubin}, \bibinfo{person}{Mahesh Ravishankar}, \bibinfo{person}{Hariharan
  Sandanagobalane}, {and} \bibinfo{person}{Vinod Grover}.}
  \bibinfo{year}{2018}\natexlab{}.
\newblock \showarticletitle{Diesel: {DSL} for linear algebra and neural net
  computations on GPUs}. In \bibinfo{booktitle}{\emph{Proceedings of the 2nd
  {ACM} {SIGPLAN} International Workshop on Machine Learning and Programming
  Languages, MAPL@PLDI 2018, Philadelphia, PA, USA, June 18-22, 2018}}.
  \bibinfo{pages}{42--51}.
\newblock
\urldef\tempurl%
\url{https://doi.org/10.1145/3211346.3211354}
\showDOI{\tempurl}


\bibitem[\protect\citeauthoryear{Hagedorn, Stoltzfus, Steuwer, Gorlatch, and
  Dubach}{Hagedorn et~al\mbox{.}}{2018}]%
        {DBLP:conf/cgo/HagedornSSGD18}
\bibfield{author}{\bibinfo{person}{Bastian Hagedorn}, \bibinfo{person}{Larisa
  Stoltzfus}, \bibinfo{person}{Michel Steuwer}, \bibinfo{person}{Sergei
  Gorlatch}, {and} \bibinfo{person}{Christophe Dubach}.}
  \bibinfo{year}{2018}\natexlab{}.
\newblock \showarticletitle{High performance stencil code generation with
  lift}. In \bibinfo{booktitle}{\emph{Proceedings of the 2018 International
  Symposium on Code Generation and Optimization, {CGO} 2018, V{\"{o}}sendorf /
  Vienna, Austria, February 24-28, 2018}}. \bibinfo{pages}{100--112}.
\newblock
\urldef\tempurl%
\url{https://doi.org/10.1145/3168824}
\showDOI{\tempurl}


\bibitem[\protect\citeauthoryear{Hartono, Norris, and Sadayappan}{Hartono
  et~al\mbox{.}}{2009}]%
        {DBLP:conf/ipps/HartonoNS09}
\bibfield{author}{\bibinfo{person}{Albert Hartono}, \bibinfo{person}{Boyana
  Norris}, {and} \bibinfo{person}{Ponnuswamy Sadayappan}.}
  \bibinfo{year}{2009}\natexlab{}.
\newblock \showarticletitle{Annotation-based empirical performance tuning using
  Orio}. In \bibinfo{booktitle}{\emph{23rd {IEEE} International Symposium on
  Parallel and Distributed Processing, {IPDPS} 2009, Rome, Italy, May 23-29,
  2009}}. \bibinfo{pages}{1--11}.
\newblock
\urldef\tempurl%
\url{https://doi.org/10.1109/IPDPS.2009.5161004}
\showDOI{\tempurl}


\bibitem[\protect\citeauthoryear{Henriksen, Serup, Elsman, Henglein, and
  Oancea}{Henriksen et~al\mbox{.}}{2017}]%
        {DBLP:conf/pldi/HenriksenSEHO17}
\bibfield{author}{\bibinfo{person}{Troels Henriksen}, \bibinfo{person}{Niels
  G.~W. Serup}, \bibinfo{person}{Martin Elsman}, \bibinfo{person}{Fritz
  Henglein}, {and} \bibinfo{person}{Cosmin~E. Oancea}.}
  \bibinfo{year}{2017}\natexlab{}.
\newblock \showarticletitle{Futhark: purely functional GPU-programming with
  nested parallelism and in-place array updates}. In
  \bibinfo{booktitle}{\emph{Proceedings of the 38th {ACM} {SIGPLAN} Conference
  on Programming Language Design and Implementation, {PLDI} 2017, Barcelona,
  Spain, June 18-23, 2017}}. \bibinfo{pages}{556--571}.
\newblock
\urldef\tempurl%
\url{https://doi.org/10.1145/3062341.3062354}
\showDOI{\tempurl}


\bibitem[\protect\citeauthoryear{Kruse and Finkel}{Kruse and Finkel}{2018}]%
        {DBLP:conf/iwomp/KruseF18}
\bibfield{author}{\bibinfo{person}{Michael Kruse} {and} \bibinfo{person}{Hal
  Finkel}.} \bibinfo{year}{2018}\natexlab{}.
\newblock \showarticletitle{A Proposal for Loop-Transformation Pragmas}. In
  \bibinfo{booktitle}{\emph{Evolving OpenMP for Evolving Architectures - 14th
  International Workshop on OpenMP, {IWOMP} 2018, Barcelona, Spain, September
  26-28, 2018, Proceedings}}. \bibinfo{pages}{37--52}.
\newblock
\urldef\tempurl%
\url{https://doi.org/10.1007/978-3-319-98521-3\_3}
\showDOI{\tempurl}


\bibitem[\protect\citeauthoryear{Phothilimthana, Elliott, Wang, Jangda,
  Hagedorn, Barthels, Kaufman, Grover, Torlak, and Bod{\'{\i}}k}{Phothilimthana
  et~al\mbox{.}}{2019}]%
        {DBLP:conf/asplos/PhothilimthanaE19}
\bibfield{author}{\bibinfo{person}{Phitchaya~Mangpo Phothilimthana},
  \bibinfo{person}{Archibald~Samuel Elliott}, \bibinfo{person}{An Wang},
  \bibinfo{person}{Abhinav Jangda}, \bibinfo{person}{Bastian Hagedorn},
  \bibinfo{person}{Henrik Barthels}, \bibinfo{person}{Samuel~J. Kaufman},
  \bibinfo{person}{Vinod Grover}, \bibinfo{person}{Emina Torlak}, {and}
  \bibinfo{person}{Rastislav Bod{\'{\i}}k}.} \bibinfo{year}{2019}\natexlab{}.
\newblock \showarticletitle{Swizzle Inventor: Data Movement Synthesis for {GPU}
  Kernels}. In \bibinfo{booktitle}{\emph{Proceedings of the Twenty-Fourth
  International Conference on Architectural Support for Programming Languages
  and Operating Systems, {ASPLOS} 2019, Providence, RI, USA, April 13-17,
  2019}}. \bibinfo{pages}{65--78}.
\newblock
\urldef\tempurl%
\url{https://doi.org/10.1145/3297858.3304059}
\showDOI{\tempurl}


\bibitem[\protect\citeauthoryear{Ragan{-}Kelley, Adams, Paris, Levoy,
  Amarasinghe, and Durand}{Ragan{-}Kelley et~al\mbox{.}}{2012}]%
        {DBLP:journals/tog/Ragan-KelleyAPLAD12}
\bibfield{author}{\bibinfo{person}{Jonathan Ragan{-}Kelley},
  \bibinfo{person}{Andrew Adams}, \bibinfo{person}{Sylvain Paris},
  \bibinfo{person}{Marc Levoy}, \bibinfo{person}{Saman~P. Amarasinghe}, {and}
  \bibinfo{person}{Fr{\'{e}}do Durand}.} \bibinfo{year}{2012}\natexlab{}.
\newblock \showarticletitle{Decoupling algorithms from schedules for easy
  optimization of image processing pipelines}.
\newblock \bibinfo{journal}{\emph{{ACM} Trans. Graph.}} \bibinfo{volume}{31},
  \bibinfo{number}{4} (\bibinfo{year}{2012}), \bibinfo{pages}{32:1--32:12}.
\newblock
\urldef\tempurl%
\url{https://doi.org/10.1145/2185520.2185528}
\showDOI{\tempurl}


\bibitem[\protect\citeauthoryear{Ragan{-}Kelley, Barnes, Adams, Paris, Durand,
  and Amarasinghe}{Ragan{-}Kelley et~al\mbox{.}}{2013}]%
        {DBLP:conf/pldi/Ragan-KelleyBAPDA13}
\bibfield{author}{\bibinfo{person}{Jonathan Ragan{-}Kelley},
  \bibinfo{person}{Connelly Barnes}, \bibinfo{person}{Andrew Adams},
  \bibinfo{person}{Sylvain Paris}, \bibinfo{person}{Fr{\'{e}}do Durand}, {and}
  \bibinfo{person}{Saman~P. Amarasinghe}.} \bibinfo{year}{2013}\natexlab{}.
\newblock \showarticletitle{Halide: a language and compiler for optimizing
  parallelism, locality, and recomputation in image processing pipelines}. In
  \bibinfo{booktitle}{\emph{{ACM} {SIGPLAN} Conference on Programming Language
  Design and Implementation, {PLDI} '13, Seattle, WA, USA, June 16-19, 2013}}.
  \bibinfo{pages}{519--530}.
\newblock
\urldef\tempurl%
\url{https://doi.org/10.1145/2491956.2462176}
\showDOI{\tempurl}


\bibitem[\protect\citeauthoryear{Rasch, Haidl, and Gorlatch}{Rasch
  et~al\mbox{.}}{2017}]%
        {DBLP:conf/hpcc/RaschHG17}
\bibfield{author}{\bibinfo{person}{Ari Rasch}, \bibinfo{person}{Michael Haidl},
  {and} \bibinfo{person}{Sergei Gorlatch}.} \bibinfo{year}{2017}\natexlab{}.
\newblock \showarticletitle{{ATF:} {A} Generic Auto-Tuning Framework}. In
  \bibinfo{booktitle}{\emph{19th {IEEE} International Conference on High
  Performance Computing and Communications; 15th {IEEE} International
  Conference on Smart City; 3rd {IEEE} International Conference on Data Science
  and Systems, HPCC/SmartCity/DSS 2017, Bangkok, Thailand, December 18-20,
  2017}}. \bibinfo{pages}{64--71}.
\newblock
\urldef\tempurl%
\url{https://doi.org/10.1109/HPCC-SmartCity-DSS.2017.9}
\showDOI{\tempurl}


\bibitem[\protect\citeauthoryear{Rasch, Schulze, and Gorlatch}{Rasch
  et~al\mbox{.}}{2019}]%
        {DBLP:conf/iwocl/RaschSG19}
\bibfield{author}{\bibinfo{person}{Ari Rasch}, \bibinfo{person}{Richard
  Schulze}, {and} \bibinfo{person}{Sergei Gorlatch}.}
  \bibinfo{year}{2019}\natexlab{}.
\newblock \showarticletitle{Developing High-Performance, Portable OpenCL Code
  via Multi-Dimensional Homomorphisms}. In
  \bibinfo{booktitle}{\emph{Proceedings of the International Workshop on
  OpenCL, {IWOCL} 2019, Boston, MA, USA, May 13-15, 2019.}}
  \bibinfo{pages}{4:1}.
\newblock
\urldef\tempurl%
\url{https://doi.org/10.1145/3318170.3318171}
\showDOI{\tempurl}


\bibitem[\protect\citeauthoryear{Steuwer, Remmelg, and Dubach}{Steuwer
  et~al\mbox{.}}{2017}]%
        {DBLP:conf/cgo/SteuwerRD17}
\bibfield{author}{\bibinfo{person}{Michel Steuwer}, \bibinfo{person}{Toomas
  Remmelg}, {and} \bibinfo{person}{Christophe Dubach}.}
  \bibinfo{year}{2017}\natexlab{}.
\newblock \showarticletitle{Lift: a functional data-parallel {IR} for
  high-performance {GPU} code generation}. In
  \bibinfo{booktitle}{\emph{Proceedings of the 2017 International Symposium on
  Code Generation and Optimization, {CGO} 2017, Austin, TX, USA, February 4-8,
  2017}}. \bibinfo{pages}{74--85}.
\newblock


\bibitem[\protect\citeauthoryear{van~den Brand, van Deursen, Heering, de~Jong,
  de~Jonge, Kuipers, Klint, Moonen, Olivier, Scheerder, Vinju, Visser, and
  Visser}{van~den Brand et~al\mbox{.}}{2001}]%
        {DBLP:conf/cc/BrandDHJJKKMOSVVV01}
\bibfield{author}{\bibinfo{person}{Mark van~den Brand}, \bibinfo{person}{Arie
  van Deursen}, \bibinfo{person}{Jan Heering}, \bibinfo{person}{H.~A. de Jong},
  \bibinfo{person}{Merijn de Jonge}, \bibinfo{person}{Tobias Kuipers},
  \bibinfo{person}{Paul Klint}, \bibinfo{person}{Leon Moonen},
  \bibinfo{person}{Pieter~A. Olivier}, \bibinfo{person}{Jeroen Scheerder},
  \bibinfo{person}{Jurgen~J. Vinju}, \bibinfo{person}{Eelco Visser}, {and}
  \bibinfo{person}{Joost Visser}.} \bibinfo{year}{2001}\natexlab{}.
\newblock \showarticletitle{The {ASF+SDF} Meta-environment: {A} Component-Based
  Language Development Environment}. In \bibinfo{booktitle}{\emph{Compiler
  Construction, 10th International Conference, {CC} 2001 Held as Part of the
  Joint European Conferences on Theory and Practice of Software, {ETAPS} 2001
  Genova, Italy, April 2-6, 2001, Proceedings}}. \bibinfo{pages}{365--370}.
\newblock
\urldef\tempurl%
\url{https://doi.org/10.1007/3-540-45306-7\_26}
\showDOI{\tempurl}


\bibitem[\protect\citeauthoryear{Vasilache, Zinenko, Theodoridis, Goyal,
  DeVito, Moses, Verdoolaege, Adams, and Cohen}{Vasilache
  et~al\mbox{.}}{2018}]%
        {DBLP:journals/corr/abs-1802-04730}
\bibfield{author}{\bibinfo{person}{Nicolas Vasilache},
  \bibinfo{person}{Oleksandr Zinenko}, \bibinfo{person}{Theodoros Theodoridis},
  \bibinfo{person}{Priya Goyal}, \bibinfo{person}{Zachary DeVito},
  \bibinfo{person}{William~S. Moses}, \bibinfo{person}{Sven Verdoolaege},
  \bibinfo{person}{Andrew Adams}, {and} \bibinfo{person}{Albert Cohen}.}
  \bibinfo{year}{2018}\natexlab{}.
\newblock \showarticletitle{Tensor Comprehensions: Framework-Agnostic
  High-Performance Machine Learning Abstractions}.
\newblock \bibinfo{journal}{\emph{CoRR}}  \bibinfo{volume}{abs/1802.04730}
  (\bibinfo{year}{2018}).
\newblock
\showeprint[arxiv]{1802.04730}
\urldef\tempurl%
\url{http://arxiv.org/abs/1802.04730}
\showURL{%
\tempurl}


\bibitem[\protect\citeauthoryear{Verdoolaege, Guelton, Grosser, and
  Cohen}{Verdoolaege et~al\mbox{.}}{2014}]%
        {Verdoolaege2014impact}
\bibfield{author}{\bibinfo{person}{Sven Verdoolaege}, \bibinfo{person}{Serge
  Guelton}, \bibinfo{person}{Tobias Grosser}, {and} \bibinfo{person}{Albert
  Cohen}.} \bibinfo{year}{2014}\natexlab{}.
\newblock \showarticletitle{Schedule Trees}. In
  \bibinfo{booktitle}{\emph{Proceedings of the 4th International Workshop on
  Polyhedral Compilation Techniques}},
  \bibfield{editor}{\bibinfo{person}{Sanjay Rajopadhye} {and}
  \bibinfo{person}{Sven Verdoolaege}} (Eds.). \bibinfo{address}{Vienna,
  Austria}.
\newblock


\bibitem[\protect\citeauthoryear{Verdoolaege, Juega, Cohen, G{\'{o}}mez,
  Tenllado, and Catthoor}{Verdoolaege et~al\mbox{.}}{2013}]%
        {DBLP:journals/taco/VerdoolaegeJCGTC13}
\bibfield{author}{\bibinfo{person}{Sven Verdoolaege},
  \bibinfo{person}{Juan~Carlos Juega}, \bibinfo{person}{Albert Cohen},
  \bibinfo{person}{Jos{\'{e}}~Ignacio G{\'{o}}mez}, \bibinfo{person}{Christian
  Tenllado}, {and} \bibinfo{person}{Francky Catthoor}.}
  \bibinfo{year}{2013}\natexlab{}.
\newblock \showarticletitle{Polyhedral parallel code generation for {CUDA}}.
\newblock \bibinfo{journal}{\emph{{TACO}}} \bibinfo{volume}{9},
  \bibinfo{number}{4} (\bibinfo{year}{2013}), \bibinfo{pages}{54:1--54:23}.
\newblock
\urldef\tempurl%
\url{https://doi.org/10.1145/2400682.2400713}
\showDOI{\tempurl}


\bibitem[\protect\citeauthoryear{Visser, Benaissa, and Tolmach}{Visser
  et~al\mbox{.}}{1998}]%
        {DBLP:conf/icfp/VisserBT98}
\bibfield{author}{\bibinfo{person}{Eelco Visser},
  \bibinfo{person}{Zine{-}El{-}Abidine Benaissa}, {and}
  \bibinfo{person}{Andrew~P. Tolmach}.} \bibinfo{year}{1998}\natexlab{}.
\newblock \showarticletitle{Building Program Optimizers with Rewriting
  Strategies}. In \bibinfo{booktitle}{\emph{Proceedings of the third {ACM}
  {SIGPLAN} International Conference on Functional Programming {(ICFP} '98),
  Baltimore, Maryland, USA, September 27-29, 1998.}} \bibinfo{pages}{13--26}.
\newblock
\urldef\tempurl%
\url{https://doi.org/10.1145/289423.289425}
\showDOI{\tempurl}


\bibitem[\protect\citeauthoryear{Zhang, Yang, Baghdadi, Kamil, Shun, and
  Amarasinghe}{Zhang et~al\mbox{.}}{2018}]%
        {DBLP:journals/pacmpl/ZhangYBKSA18}
\bibfield{author}{\bibinfo{person}{Yunming Zhang}, \bibinfo{person}{Mengjiao
  Yang}, \bibinfo{person}{Riyadh Baghdadi}, \bibinfo{person}{Shoaib Kamil},
  \bibinfo{person}{Julian Shun}, {and} \bibinfo{person}{Saman~P. Amarasinghe}.}
  \bibinfo{year}{2018}\natexlab{}.
\newblock \showarticletitle{GraphIt: a high-performance graph {DSL}}.
\newblock \bibinfo{journal}{\emph{{PACMPL}}} \bibinfo{volume}{2},
  \bibinfo{number}{{OOPSLA}} (\bibinfo{year}{2018}),
  \bibinfo{pages}{121:1--121:30}.
\newblock
\urldef\tempurl%
\url{https://doi.org/10.1145/3276491}
\showDOI{\tempurl}


\end{thebibliography}
